\documentclass[12pt]{JHEP3}
\usepackage{graphicx}
\usepackage{dcolumn}  
\usepackage{bm}    
\usepackage{amssymb} 
\usepackage{amsmath,bm}
\usepackage{amsfonts}    
\usepackage{slashed}  
\title{Higher spin quasinormal modes and one-loop determinants  in the BTZ black hole }
\author{Shouvik Datta,  Justin R. David   \\
 Centre for High Energy Physics,
Indian Institute of Science,\\ C.V. Raman Avenue, Bangalore 560012, India. \\
\email{shouvik, justin@cts.iisc.ernet.in}\\
}

\abstract{ 
We solve the wave equations of arbitrary integer spin fields in the 
BTZ black hole background and obtain exact expressions for their  quasinormal modes. 
We show that these quasinormal modes precisely agree with the location of the  poles of the 
corresponding two point function in the dual conformal field theory as predicted 
by the AdS/CFT correspondence.  
We then use these quasinormal modes to construct the one-loop
determinant of the higher spin  field in the thermal BTZ background.
This is shown to agree with that obtained from the  corresponding heat kernel 
constructed  recently by group theoretic methods. 
}

\begin{document}

\section{Introduction}

Among the gauge/gravity dualities 
the  $AdS_3/CFT_2$ correspondence is one of the most well studied. 
In string theory, this duality naturally occurs in the context of the D1-D5 system 
\cite{Maldacena:1998bw}. 
Till recently this example and  related systems   were the only examples with a well defined  
conformal field  theory to study the 
$AdS_3/CFT_2$ duality. The proposal put forward  
by   Gaberdiel and Gopakumar \cite{Gaberdiel:2010pz} provides a new and interesting 
example to explore this version of the gauge/gravity duality. 
In some respects this example might be more tractable since  the bulk theory  does not contain 
the full string spectrum  but a set of massless higher spin fields. 
All these examples contain interacting higher spin fields.  The D1-D5 system and its cousins
 in general contain interacting higher spin massive fields, while the $AdS_3$ duals of minimal 
models put forward by \cite{Gaberdiel:2010pz} contain higher spin massless fields.
In this paper we will focus on the more general situation when the 
higher spin field is massive.  
 $AdS_3/CFT_2$ duality relates a field of spin $s$ propagating in 
$AdS_3$ to a operator ${\cal O}$ in the dual 
conformal field theory characterized by conformal 
weights $(h_L, h_R)$ with \cite{Aharony:1999ti}
\begin{equation}
 h_R-h_L = \pm s .
\end{equation}
The mass of the propagating field $m$  is related to the conformal dimension of the operator
${\cal O}$ which is given by
\begin{equation}
 h_R+ h_L =\hat  \Delta.
\end{equation}
An immediate consequence of this correspondence is that when the CFT is at finite temperature, 
the poles in the retarded  point function of the operator ${\cal O}$ is given by 
quasinormal modes of the spin $s$ field \cite{Birmingham:2001pj,Son:2002sd,Kovtun:2005ev}. 
In $CFT_2$, two point functions of primary operators are determined entirely by conformal
invariance and one can read out the poles of the retarded Green's in the  complex frequency  plane.
These are given by
\begin{equation}
\label{qnm}
 \omega_L = k - 4\pi i T_L ( n + h_L), \qquad \omega_R = -k -4\pi i T_R ( n + h_R) . 
\end{equation}
Here $\omega$ and $k$ refers to the frequency and momentum respectively. $L, R$ are subscripts 
to denote the left and right moving poles.  $T_L, T_R$ are the left and right moving 
temperatures of the conformal field theory. 
Thus, the $AdS_3/CFT_2$ correspondence predicts  the quasinormal frequencies of fields with 
spin $s$. One of the aims of this 
paper is  to verify this  prediction  for the case of arbitrary integer spins. 

 Quasinormal modes of various fields in the black hole background 
 play an important role in the AdS/CFT  correspondence.
Their role for black holes in anti-de Sitter spaces were first investigated in   \cite{Horowitz:1999jd}.
The low lying quasinormal modes provide important information regarding 
transport properties of the dual field theory.  A recent review of quasinormal modes 
in various backgrounds, their properties and a complete list of references is \cite{Berti:2009kk}. 
In most of the situations  the complete spectrum of quasinormal modes can only 
be obtained numerically.   Integrability of the string propagation in the 
BTZ background \cite{David:2011iy}  suggests that it might be possible to obtain 
the quasinormal modes of arbitrary spin fields exactly. 
Indeed  for the case of  spins $s=0$,  $s= 1/2$ and $s=1$  the corresponding wave equation were 
solved exactly \cite{Birmingham:2001hc, Das:1999pt}. This  confirmed the prediction 
of the AdS/CFT duality given in  (\ref{qnm}) \cite{Birmingham:2001pj}.  
These reasons and the simplicity of the 
expressions for the quasinomal spectrum in (\ref{qnm})  are the motivations 
for deriving the quasinormal modes for higher spins. 
Another interesting property of quasinormal modes discovered recently 
is that the 1-loop determinant of the corresponding
field can be constructed by considering suitable products of the quasinormal modes \cite{Denef:2009kn}. 
The one-loop  determinants for  arbitrary spin fields in thermal $AdS_3$ was
recently constructed in \cite{David:2009xg} using group theoretic methods.  In this paper we
show that these one-loop determinants can be written as product 
over higher spin quasinormal modes.

The organization of the paper is as follows. 
We will begin with some preliminaries where will we introduce the
BTZ background and its properties, the  equations satisfied by 
massive higher spin fields and  the AdS/CFT  prediction for the 
quasinormal modes from the CFT.
In   section 3, as a simple demonstration of the general method we  develop in this paper
we analyze the case of spin 2.  We show the 
 equations of motion for the spin 
2 field can be simplified and its solutions can be found in
closed form as hypergeometric functions. We then extract the quasinormal modes
and show it agrees with that given in (\ref{qnm}). 
In section 4 we generalize this analysis for  integer  higher spins.  It turns out 
that the wave equations for these cases also can be solved in terms of hypergeometric functions. 
In section 5, we write down  the one-loop
determinant  for the higher spin field in terms of products over the
 corresponding quasinormal modes and show that it 
agrees with that evaluated by group theoretic methods.  Section 6 contains our conclusions.
Appendix A  and Appendix B contains the details of the quasinormal modes for spin 1 and spin 3 respectively. 
Appendix C contains  the proofs of identities which are  required  in our analysis 
for the solutions of higher spin wave equations  in section 4. 

\section{Preliminaries}

\def\sg{\sqrt{-g}}
\def\pd{\partial}

In this section we will describe the geometry of the BTZ black hole and its properties.
This will help us introduce the conventions and notations we will use in this paper. 
We will also describe the key properties of the BTZ background  which will 
enable us to simplify the equations of motion for the higher spin fields in this background. 
We then review the properties of the  higher spin field equations.  
Finally we  recall  the AdS/CFT dictionary for these fields and write down  the conformal dimensions of the 
operator corresponding to these fields and their two point functions. 
We then extract the poles of the these two point functions. 
By the AdS/CFT correspondence, these poles must coincide with 
the quasinormal modes of the higher spin field in the bulk.  In the subsequent  sections we will focus on 
extracting these quasinormal modes by explicitly solving the equations of motion of the higher spin
fields and we will demonstrate the 
quasinormal modes do  indeed  coincide with these poles. 

\subsection{The black hole geometry}

The BTZ black hole  is a solution of Einstein's gravity with a 
negative cosmological constant in (2+1) dimensions \cite{Banados:1992wn}.
 In general it describes a spinning  black hole 
which asymptotes to $AdS_3$.  Its metric is conventionally written as 
\begin{eqnarray}
\label{metconv}
 ds^2 &=&  -\frac{\Delta^2}{ r^2} dt^2 + \frac{r^2}{\Delta^2} dr^2 + 
r^2 \left( d\phi - \frac{r_+r_-}{r^2} dt\right)^2, \\ \nonumber
\Delta^2 & = & ( r^2 - r_+^2) ( r^2 - r_-^2).  
\end{eqnarray}
Here $r_+$ and $r_-$ are the radii of the inner and outer horizons respectively,  
$r$ is the radial distance and $t$ labels the time. The angular coordinate
$\phi$ has the period of $2\pi$.  
The radii $r_+$ and $r_-$ are related to 
  $M$, the mass of black hole and  $J$, its  angular momentum by the following 
expression
\begin{align}
r_\pm = \frac{Ml^2}{2} \left(   1\pm \sqrt{1-\frac{J^2}{M^2 l^2}} \right).
\end{align}
where $l$ is the radius of anti-de Sitter space. From this point onwards we will set the radius
of $AdS_3$ to unity. 
 The left and right temperatures are defined as 
\begin{equation}
 T_L = \frac{1}{2\pi} ( r_+ - r_-) , \qquad T_R = \frac{1}{2\pi} ( r_+ + r_-) .
\end{equation}
The two temperatures  capture the fact this  system has
two intrinsic thermodyamical  variables: the thermal temperature and the angular potential. 
From the boundary conformal field theory point of view they define the two temperatures of the 
left movers and right movers. 

A convenient coordinate system for our analysis was discovered by \cite{Birmingham:2001pj}. We first define the coordinates
\begin{align}
\label{xpxm}
z = \tanh^2 \xi  &= \frac{r^2 - r_+^2}{r^2 - r_-^2}, \\ \nonumber
 x^+ = r_+ t - r_- \phi,& \qquad x^- = r_+ \phi - r_- t.
\end{align}
Note that in these coordinates, the range of $r$ from $r_+$ to $\infty$ is mapped to 
$z=0$ or $\xi =0$ to $z=1$ or $\xi = \infty$ respectively. 
In these coordinates, the BTZ metric given in (\ref{metconv})  reduces to  the following diagonal metric
\begin{align}
\label{diamet}
ds^2 &= d\xi ^2 - \sinh^2 \xi \, dx_+^2 + \cosh^2 \xi \, dx_-^2 .
\end{align}
This form of the metric will prove to be useful in our calculations and we will briefly list various
properties of this metric which will be repeatedly used. 
The non-vanishing Christoffel  symbols of the metric in (\ref{diamet}) are  given by
\begin{eqnarray}
\begin{aligned}
\label{chris}
 &\Gamma_{++}^\xi = \cosh\xi \sinh\xi =\frac{\sqrt{z}}{1-z}, \quad &&\Gamma^\xi_{--} = - \cosh\xi \sinh \xi = -\frac{\sqrt{z}}{1-z} , \\ \nonumber
 &\Gamma^{+}_{+\xi} = \coth \xi=\frac{1}{\sqrt{z}}, \quad &&\Gamma^{-}_{- \xi} = \tanh\xi=\sqrt{z} \ . 
\end{aligned}
\end{eqnarray}
The metric and its Christoffel symbols obey the following identities
which will be useful in simplifying the higher spin equations in the next sections 
\begin{eqnarray}
\sg &=&  \cosh \xi \sinh \xi = \frac{\sqrt{z}}{1-z}\ ,\\
\frac{g_{++}}{\sg} &=& - \tanh \xi =-\sqrt{z} \ , \\
\frac{g_{--}}{\sg} &=&  \coth \xi =\frac{1}{\sqrt{z}} \ , \\
 \label{gdg} \frac{1}{\sg} \pd _\mu (\sg g^{\mu \nu} \Gamma ^{\sigma}_{\nu \rho}) 
&= & \label{c0}
\frac{1}{\sg} \pd _\xi (\sg g^{\xi\xi} \Gamma ^{\sigma}_{\xi \rho}) = 2 \hat\delta _{\rho}^{\sigma}  \\
\label{c1} g^{\pm \pm}\Gamma^\xi _{\pm \pm} + \Gamma ^\pm _{\xi\pm} &=&  0  \ , \\
g^{++}\Gamma^\xi _{++} &=&  - \coth \xi =-\frac{1}{\sqrt{z}}  \ , \\
g^{--}\Gamma^\xi _{--} &=&  - \tanh \xi =-\sqrt{z}  \ .
\end{eqnarray}
where $\hat \delta_{\rho\sigma}$ is defined as
\begin{equation}
\hat\delta_{\rho}^{\sigma} = 
\begin{cases}
1 & \text{for $\rho,\sigma = \pm$ and $\rho=\sigma$,}\\
0 & \text{otherwise.}
\end{cases}
\end{equation}

The BTZ black hole is obtained by 
identifications of $AdS_3$ \cite{Banados:1992gq}. 
Thus it is locally $AdS_3$ and therefore its curvature  obey the following relations
\begin{eqnarray}
\label{local}
R_{\alpha\beta\gamma\delta} &=& g_{\alpha\delta}g_{\beta\gamma}-g_{\alpha\gamma}g_{\beta\delta}, \\
R_{\mu\nu} &=& -2g_{\mu\nu}, \qquad  G_{\mu\nu} =  4 g_{\mu\nu}.
\end{eqnarray}
In 3 dimensions, the Riemann tensor further obeys the following relation
\begin{eqnarray}
\label{3dreim}
 R_{\alpha\beta\gamma\delta} &=& \epsilon_{\alpha\beta\rho}\epsilon_{\gamma\delta\sigma}( 
R^{\rho\sigma} -  \frac{1}{2}R g^{\rho\sigma} ), \\ \nonumber
&=& \epsilon_{\alpha\beta\rho}\epsilon_{\gamma\delta\sigma}G^{\rho\sigma} .
\end{eqnarray}
Here  $G^{\rho\sigma}$ is the Einstein tensor and the 
epsilon tensor is defined as
\begin{equation}
\epsilon^{\alpha\beta\gamma}=\frac{\widetilde{\epsilon}^{\ \alpha\beta\gamma}}{\sg}, \qquad \widetilde{\epsilon} ^{\ +\xi -} = 1.
\end{equation}
where, \( \widetilde{\epsilon}^{\alpha\beta\gamma}\) is the completely antisymmetric Levi-Civita symbol. Finally, we also need the fact that the  epsilon tensor in 3 dimensions satisfies the relation
\begin{equation}
\label{epident}
 \epsilon^{\ \ \, \alpha}_{\beta\rho} \epsilon_{\alpha\delta\sigma} = - 
( g_{\beta\delta}g_{\rho\sigma} - g_{\beta\sigma}g_{\rho\delta} ) .
\end{equation}

\subsection{The description of higher spin fields}

\def\NU{\nu_2\nu_3\cdots\nu_s}
\def\nn{\nonumber}

Massive integer spin $s$ fields in $AdS$ spaces are realized by 
totally symmetric tensors of rank $s$ satisfying  
the following equations \cite{Metsaev:2003cu,Buchbinder:2006ge}
\begin{eqnarray}
\label{eom-s} ( \nabla^2  -m^2_s ) \Phi_{\mu_1\mu_2\mu_2 \cdots \mu_s} &=&0,  \\
 \nabla^\mu \Phi_{\mu \mu_2 \cdots \mu_s} &=& 0, \\
 g^{\mu\nu} \Phi_{\mu\nu \mu_3 \cdots \mu_s} &=& 0.
\end{eqnarray}
Here  $\Phi_{\mu_1 \mu_2 \cdots \mu_s}$ is a totally symmetric rank $s$ tensor and for $AdS_3$ the 
mass is given by
\begin{equation}\label{defms}
m_s^2 = s(s-3) +M^2 .
\end{equation}
The first term is the  natural mass that exists due to the curvature of $AdS_3$. Note that it also exists when 
$M=0$. 

A fact that will be crucial in our analysis is that for  massive spin $s$ field in $AdS_3$ the set of 
equations in (\ref{eom-s}) is equivalent the following first order equation
\cite{Tyutin:1997yn}
\begin{equation}\label{spinCS}
\epsilon _{\mu}^{\ \alpha\beta} \nabla _\alpha \Phi _{\beta\nu_2\nu_3\cdots\nu_{s}}=-m \Phi _{\mu\nu_2\nu_3\cdots\nu_{s}}.
\end{equation}
Such an equation is familiar  for the massive  spin-1 field in $AdS_3$. 
It is well known that for the spin-1 field, the massive Chern-Simons equation is 
equivalent to the massive  Maxwell field.  We will now demonstrate that the equation in 
(\ref{spinCS}) in a locally $AdS_3$ space is  equivalent to the second order equation of motion
(\ref{eom-s}) and the gauge constraint. 
To verify the gauge constraint we take the covariant derivative on both sides of (\ref{spinCS}). 
This results in 
\begin{eqnarray}
-m \nabla^\mu \Phi_{\mu \nu_2 \nu_3 \cdots \nu_s} &=&  
\epsilon^{\mu\alpha\beta}\nabla_\mu\nabla_\alpha \Phi_{\beta\nu_2 \nu_3 \cdots \nu_s} 
=  \frac{1}{2} \epsilon^{\mu\alpha\beta}[\nabla_\mu, \nabla_\alpha] 
\Phi_{\beta\nu_2\nu_3\cdots \nu_s} \\ \nonumber
&=& \epsilon^{\mu\alpha\beta} g^{\rho\rho'} R_{\beta\rho\mu\alpha} \Phi_{\rho'\nu_2\nu_3 \cdots \nu_s}
+ \epsilon^{\mu\alpha\beta}g^{\rho\rho'}  R_{\nu_2 \rho\mu\alpha}\Phi_{\beta\nu_2\rho'\cdots \nu_s}
+ \cdots \\ \nonumber
&=&\epsilon^{\mu\alpha\beta} \epsilon_{\beta\rho\delta} \epsilon_{\mu\alpha\delta'} 
G^{\delta\delta'} g^{\rho\rho'} \Phi_{\rho'\nu_2\nu_3 \cdots \nu_s}
+ \epsilon^{\mu\alpha\beta} \epsilon_{\nu_2\rho\delta} \epsilon_{\mu\alpha\delta'} 
G^{\delta\delta'} g^{\rho\rho'} \Phi_{\beta\nu_2 \rho'\cdots \nu_s}
+ \cdots \\ \nonumber
&=& - 2 \epsilon_{\beta\rho\delta} G^{\delta\beta} g^{\rho\rho'} \Phi_{\rho'\nu_2\nu_3\cdots \nu_s}, 
-2 \epsilon_{\nu_2 \rho\delta} G^{\delta\beta} g^{\rho\rho'} \Phi_{\beta\rho'\nu_3\cdots \nu_s} + \cdots
 \\ \nonumber
&=& -8 \epsilon_{\nu_2}^{\rho'\beta} \Phi_{\beta\rho'\nu_3 \cdots \nu_s} + \cdots \\ \nonumber
&=& 0 .
\end{eqnarray}
Here we have used (\ref{3dreim}) and (\ref{epident}) to arrive at the third and fourth line of the 
above equation.   In the last but one line we have used (\ref{local}) which results from the fact the 
the space is locally $AdS_3$.  The last line results from 
 the traceless condition of the spin $s$ field. 
Thus the gauge constraint is  satisfied once the symmetric traceless 
field satisfies (\ref{spinCS}). 

We will now demonstrate that the second equation in (\ref{eom-s}) is implied once the 
the field satisfies (\ref{spinCS}). 
Substituting for $\Phi _{\beta\nu_2\nu_3\cdots\nu_{s}}$ on the left hand side of  (\ref{spinCS})
by using the equation itself we obtain
\begin{equation}
\epsilon_\mu ^{\ \alpha\beta} \epsilon _\beta ^{\ \rho\sigma} \nabla _\alpha \nabla _\rho \Phi _{\sigma\NU} =m^2 \Phi _{\mu\NU}.
\end{equation}
Using (\ref{epident}) and rearranging the equation we obtain,
\begin{eqnarray}
\label{2ndorder}
 (\nabla ^2 -m^2)  \Phi _{\mu\NU} &=&  \nabla ^\sigma \nabla _\mu  \Phi _{\sigma\NU} \nn \\
&= & g^{\sigma\rho} [\nabla_\rho,\nabla_\mu]\Phi_{\sigma\NU}  \nn \\
&=&   g^{\sigma\rho} g^{\eta\delta} (R_{\sigma\delta\rho\mu}\Phi_{\eta\NU}+R_{\nu_2\delta\rho\mu}\Phi_{\sigma\eta\nu_3\cdots\nu_s}
\nn \\
& & \qquad \qquad+R_{\nu_3\delta\rho\mu}\Phi_{\sigma\nu_2\eta\nu_4\cdots\nu_s}+\cdots+R_{\nu_s\delta\rho\mu}\Phi_{\sigma\nu_2\nu_3\cdots\nu_{s-1}\eta}) \nn \\
&=& -(s+1)\Phi_{\mu\NU} .
\end{eqnarray}
To obtain the equality on the first line we have used the gauge constraint and finally to 
obtain the last line we have 
used the form of the Riemann tensor given in (\ref{3dreim}). 
Thus we obtain the following second order equations of motion of the spin 
$s$ field 
\begin{equation}
 (\nabla ^2 -m^2 + (s+1))  \Phi _{\mu\NU} =0. \label{2ndOrder}
\end{equation}
We are now in a position to relate the mass, $m$ appearing in the first order equation (\ref{spinCS}) to the actual mass, $M$ of the higher spin field. Comparing, (\ref{2ndOrder}) with (\ref{eom-s}) we get,
\begin{equation}
\label{csmass}
m^2 = M^2 + (s-1)^2.
\end{equation}

As a consistency check we  can use the above gauge condition and the tracelessness 
condition of the spin $s$ field to show
that the the LHS of the equation (\ref{spinCS}) is symmetric  the indices $\mu$ and $\nu_2$. 
To do this  we must show 
that the anti-symmetry combination of the LHS of (\ref{spinCS}) vanishes. 
This is equivalent to showing the following vanishes
\begin{align}
\epsilon^{\mu\nu_2\rho} \epsilon_{\mu}^{\ \alpha\beta} \nabla_\alpha \Phi_{\beta\nu_2\cdots \nu_s}
&= - ( g^{\nu_2\alpha} g^{\rho\beta}  - g^{\nu_2\beta} g^{\rho\alpha} ) 
\nabla_{\alpha} \Phi_{\beta\nu_2 \nu_3 \cdots \nu_s}\nonumber \\ \nonumber
&= - g^{\rho\beta} \nabla^\alpha\Phi_{\beta\alpha\nu_3 \cdots \nu_s} +
g^{\rho\alpha} \nabla_\alpha g^{\nu_2\beta} \Phi_{\beta\nu_2 \nu_3 \cdots \nu_s}
\\ 
&= 0.
\end{align}
Here we have used (\ref{epident}) to obtain the 
first equation. Then the gauge condition and the  tracelessness condition implies that
the antisymmetric component of the LHS of (\ref{spinCS}) vanishes. 

Finally, it is instructive to count the number of degrees of freedom of the spin $s$ 
field in 3 dimensions which satisfies the gauge constraint and the 
tracelessness constraint in  (\ref{eom-s}). 
The number of independent components of a totally symmetric rank $s$ tensor 
in 3 dimensions which is traceless is given by
\begin{equation}
D(s) = \frac{( s+2)(s+1)}{2} - \frac{s( s-1)}{2} = 2s+1 .
\end{equation}
The number of independent equations in the gauge constraint is given by
\begin{equation}
C(s) = \frac{(s+1)s}{2} - \frac{(s-1)(s-2)}{2} = 2s-1.
\end{equation}
Here again the subtraction is due to the traceless condition. 
Thus the total number of degrees of freedom of a rank  $s$   symmetric traceless 
field which satisfies the gauge constraint is
\begin{equation}
D(s) - C(s) = 2 .
\end{equation}

\subsection{Higher spin fields  in  $AdS_3/CFT_2$}

We now recall the relation between a rank $s$ symmetric  tensor which satisfies 
the equations (\ref{eom-s}) in $AdS_3$ and the corresponding operator on the boundary. 
We first find the relation between the mass of the spin $s$ field and the dimension of the 
operator. For this we must investigate the behaviour of the solutions to 
(\ref{eom-s}) close to the boundary. 
Consider Euclidean $AdS_3$ in Poincare coordinates with the metric
\begin{equation}
ds^2 = \frac{1}{\hat z^2} \left( d\hat z^2 + (dx^1)^2 + ( dx^2)^2 \right) .
\end{equation}
Then the spin $s$ Laplacian can be written as \cite{Giombi:2009wh}
\begin{eqnarray}
\nabla^2 \Phi_{\mu_1 \cdots \mu_s} &=& 
  \hat z^2 \left[ \big (\partial_{\hat z}  + \frac{s-1}{\hat z} \big ) \big ( \partial_{\hat z } + \frac{s}{z} \big ) 
+ \hat z^2 \partial_i \partial_ i - s \right] \Phi_{\mu_1 \cdots \mu_s} \\ \nonumber
& & - 2 s \hat z \partial_{( \mu_1} \Phi_{\mu_2\cdots \mu_s ) \hat z }  + s( s-1) \eta_{(\mu_1\mu_2}
\Phi_{\mu_3 \cdots \mu_s ) \hat z\hat z} \\ \nonumber
& & - s ( 2s-1) \delta_{\hat z(\mu_1} \Phi_{\mu_2 \cdots \mu_s ) \hat z} 
+ 2 s \hat z \partial_\rho \delta_{\hat z( \mu_1 } \Phi_{\mu_2 \cdots \mu_s ) \rho}.
\end{eqnarray}
while the gauge condition reduces to
\begin{equation}
\label{bgc}
 ( \partial_{\hat z }- \frac{1}{\hat z})  \Phi_{\hat z \mu_2 \cdots \mu_{s}} + 
\partial_i \Phi_{i \mu_1 \cdots \mu_{s}} =0.
\end{equation}
To obtain this equation the traceless property of the spin $s$ tensor was used. 
We can now solve for the behaviour of these fields close to the boundary. 
Let us make the ansatz that
\begin{equation}
\label{bcbeh}
 \Phi_{i_1 \cdots i_s}( x, \hat z ) \sim z^\delta, \qquad \hat z \rightarrow 0.
\end{equation}
where $x$ and $i_k$ denote the  coordinates along the boundary. 
Then from the gauge condition (\ref{bgc})  we see that
\begin{equation}
 \Phi_{\hat z\mu_2\cdots \mu_{s}} \sim \hat z^{\delta +1} .
\end{equation}
This implies that 
the leading terms in the Laplacian and hence in the 
equations of motion near the boundary $\hat z\rightarrow 0$  are given by
\begin{eqnarray}
\left( \hat z ^2 ( \partial_{\hat z} + \frac{s-1}{\hat z } ) ( \partial_{\hat z} + \frac{s}{\hat z } )
- s  - m^2_s  \right) \Phi_{i_1 i_2 \cdots i_s} = 0.
\end{eqnarray}
Substituting the ansatz in (\ref{bcbeh}) we can 
easily solve for $\delta$ which is given by the solution of the following equation
\begin{eqnarray}
 ( \delta + s)( \delta + s -2) - s = m_s^2 .
\end{eqnarray}
Solving this quadratic equation one obtains the following two behaviours of the 
spin $s$ field at the boundary
\begin{eqnarray}
\label{massdim}
 \delta &=& - ( s-1) \pm \sqrt{ s+ 1 + m_s^2}, \\ \nonumber
&=& - ( s-1) \pm \sqrt{ ( s-1)^2 + M^2}, \\ \nonumber
&=& -(s-1) \pm |m| .
\end{eqnarray}
To obtain the second  line we have  substituted  the value of $m_s^2$ from (\ref{defms}). 
The last  line is obtained by using the relation (\ref{csmass}). 
It is important to note that this asymptotic analysis also holds for the case of the BTZ black hole. 
This is because asymptotically the BTZ black hole reduces to $AdS_3$ as can be seen 
from the metric in (\ref{metconv}) with the identification
\begin{equation}
 \hat z = \frac{1}{r}.
\end{equation}
The conformal dimension $\Delta$ of the dual operator 
  can then be obtained from 
the coupling of the boundary value of the spin $s$ field and  a spin $s$ current 
which is given by
\begin{equation}
 \int d^2 x J^{i_1, \cdots i_s} \Phi_{i_1, \cdots i_s} .
\end{equation}
From conformal invariance we obtain the following 
expression for the conformal dimension of the dual operator,
\begin{eqnarray}\label{conf-dim}
\hat\Delta &=&2-\delta -s = 1+  \sqrt{(s-1)^2+M^2}, \\ \nonumber
&=& 1 + |m| .
\end{eqnarray}
Here  we have chosen the negative branch of for $\delta$. 
Thus the conformal weights $(h_L, h_R$  of the dual operator 
are given by  relations
\begin{equation}
\label{cftdim}
 \hat\Delta = h_L + h_R,  \qquad h_R- h_L = \pm s .
\end{equation}
The second relation can be understood as follows. 
The rank $s$ traceless symmetric tensor has only two independent degrees of 
freedom. In terms of the conformal group, this is just the 
spin of the boundary operator, which can be $+s$ or $-s$. 
For definiteness we will assume that $m$ is positive and as we will subsequently see this
will lead to the situation $h_R -h_L =s$.

\vspace{.5cm}\noindent
{\bf The two point function and its poles}
\vspace{.5cm}

The  two point function of an conformal primary with weight $(h_L, h_R)$ 
with the conformal field theory at right and left temperatures $T_L, T_R$
is given  by
\begin{equation}
\label{corel}
{\cal G}(t, x) =  \langle {\cal O}(t, x) {\cal O}(0, 0 ) \rangle_T = 
\frac{ {\cal C}_{\cal O} }{ i^ {(2h_L + 2  h_R)} } 
\left(  \frac{ \pi T_L}{ \sinh \pi T_L x^+} \right)^{2h_L} 
 \left(  \frac{ \pi T_R}{ \sinh \pi T_R x^-} \right)^{2h_R} .
\end{equation}
where $x^\pm = t \pm x$.  The two point function is entirely determined by 
conformal invariance. We have followed the notations of \cite{Gubser:1997cm}
in writing down the above expression. 
The retarded Green's function is then given by
\begin{eqnarray}
\label{retarded}
 {\cal D}^{\rm ret} (x, x') = i \theta(t-t') \langle [{\cal O}(x), {\cal O}(x') ]\rangle_T, \\ \nonumber
\end{eqnarray}where the commutator is evaluated in the canonical ensemble.  From the definition given in (\ref{corel}) and (\ref{retarded}) we obtain the relation
\begin{eqnarray}
 {\cal D}^{\rm{ret}} (x, 0) &=& i \theta( t) \bar{\cal D} (x, 0) , \\ \nonumber
 \bar{\cal D} (x, 0)  &=& {\cal G}( t+ i\epsilon, x) - {\cal G}( t- i\epsilon)   .
\end{eqnarray}
The Fourier transform of $\bar{\cal D}(x, 0)$ was performed in \cite{Gubser:1997cm} and the result is given by
\begin{eqnarray}
\label{fttrans}
 & & \int d^2 x e^{-i p\cdot x}  \bar{\cal D} (x, 0) =  \\ \nonumber
 & & {\cal C}_{{ \cal O} } \frac{( 2\pi T_L)^{2h_L -1} ( 2\pi T_R)^{2h_R -1}}{\Gamma( 2h_L) \Gamma( 2 h_R) }
 \frac{ e ^{\beta\cdot p/2} - ( -1)^{2h_L + 2h_R} e^{-\beta\cdot p/2}}{2}   \\ \nonumber
& &\quad \times  \left|
\Gamma\left( h_L + i \frac{p_+}{2\pi T_L} \right) \Gamma\left( h_R + i \frac{p_-}{2\pi T_R} \right)
\right|^2  .
\end{eqnarray}
here $ p\cdot x = \omega t - k x$, $p_{\pm} = \frac{1}{2}( \omega \mp k )$. 
From this we see that the poles of the retarded Green's function are the poles that lie
in the lower half plane. The two set of poles are given by
\begin{equation}\label{polesl}
\begin{aligned}
\omega _L &=  k+2\pi T_L(n+h_L), \\
&=  k-2\pi iT_L(2n + \hat \Delta -s),
\end{aligned}\qquad
\begin{aligned}
\omega _R &= -k+2\pi T_R(n+h_R), \\ 
&= -k-2\pi iT_R (2n+\hat \Delta +s).
\end{aligned}
\end{equation}
To obtain the second line from the first we have used the relations in (\ref{cftdim}) with the positive 
sign for $s$.  Note that the function in (\ref{fttrans}) has poles also in the upper half plane. 
The reason these poles are selected out in the retarded Green's function is that the response has 
to die off as $t\rightarrow+ \infty$. Therefore the pole has to lie on the lower half plane. 
For spin $s=0, 1$ and $1/2$ it has been shown that these poles coincide  with the
quasinormal modes of the corresponding spin field in the BTZ background \cite{Birmingham:2001pj}. 
One of the aims of this paper is to generalize this result for the case of arbitrary spins.

In the next section, as a simple demonstration of the  general method for the case 
of arbitrary integer spins, we  will solve the wave equations of the massive spin 2 field in the 
BTZ background and extract out the quasinormal modes. We will demonstrate that the poles
coincide with that given in (\ref{polesl}). 
The experience gained by this analysis will enable us to generalize the result 
for the spin $s$ case.

\def\csch{\text{cosech}}
\def\sech{\text{sech}}
\def\a{\alpha}
\def\b{\beta}
\def\nn{\nonumber}

\section{The spin 2 case}

In this section we will  evaluate the quasinormal modes of the $s=2$ field. 
This will involve reduction of the spin $s$ Laplacian to a scalar Laplacian by taking appropriate linear 
combinations of the various components of the spin 2 field. 
The spin 2 case  will serve as an 
demonstration of the method we will develop for 
determining the quasinormal modes for fields of arbitrary spin  \(s\). 
We begin by writing down the first order equations 
(\ref{spinCS}) satisfied by the spin 2 field 
\begin{equation}\label{CS-2}
\epsilon_\mu ^{\ \alpha \beta} \nabla _\alpha h_{\beta \nu} = -m h_{\mu \nu}.
\end{equation}
An important fact that we will use for our analysis is that near the boundary 
components which have the radial coordinate that is $h_{\xi+}, h_{\xi -}, h_{\xi\xi}$ are 
all determined from the components along the boundary and are suppressed. 
This is clear from our analysis  around (\ref{bgc}).  quasinormal modes are obtained
by imposing vanishing Dirichlet boundary conditions which demand the function to vanish 
at the boundary,  Thus it is sufficient to examine the equations satisfied
by the components $h_{++}, h_{--}, h_{+-}$. 
From (\ref{CS-2}) we obtain the following equations satisfied by these components
\begin{align}
\label{g1} \frac{g_{++}}{\sg} (\nabla_\xi h_{-+} -\nabla_- h_{\xi +}) =  -m  h_{++},\\
\label{g2} \frac{g_{--}}{\sg} (\nabla_+ h_{\xi-} - \nabla_\xi h_{+-}) =  -m  h_{--}, \\
\label{g3} \frac{g_{++}}{\sg} (\nabla_\xi h_{--} - \nabla_- h_{\xi-}) =  -m  h_{+-}, \\
\label{g4} \frac{g_{--}}{\sg} (\nabla_+ h_{\xi+} -\nabla_\xi h_{++}) =  -m  h_{-+}. 
\end{align}
The spin-2 field is also traceless, explicitly this condition is expressed by the following equation
\begin{eqnarray}
\label{tr-g}
 h_{\xi\xi} - \csch ^2 \xi \ h_{++} + \sech ^2 \xi \ h_{--} = 0 .
\end{eqnarray}

\subsection{Reduction of the spin-2  Laplacian to the scalar Laplacian}

The first step in the analysis is to show that  the spin 2 Laplacian 
on the components reduces $h_{++}, \ h_{--}$ and $ h_{+-}$ to the 
scalar Laplacian together with a mixing  `mass matrix'. 
We start by analyzing the action of the Laplacian on \(h_{\mu\nu}\)
\begin{eqnarray}
\label{lapspin2}
\nabla ^2 h _{\mu \nu}&=& \Delta h_{\mu\nu} - \frac{1}{\sg}\pd _\alpha (\sg g^{\a\b} \Gamma ^\sigma_{\b\mu} ) h_{\sigma\nu} -  g^{\a\b} \Gamma ^\sigma_{\b\mu}  \pd _\a h_{\sigma\nu} \nn \\
& & \qquad  \ \ -  \frac{1}{\sg}\pd _\alpha (\sg g^{\a\b} \Gamma ^\sigma_{\b\nu} ) h_{\sigma\mu} -  g^{\a\b} \Gamma ^\sigma_{\b\nu}  \pd _\a h_{\sigma\mu} \nn \\
& & \qquad  \ \ -\Gamma^\rho_{\a\mu}g^{\a\b}\nabla_\b h _{\rho\nu} - \Gamma^\rho_{\a\nu}g^{\a\b}\nabla_\b h _{\rho\mu} \nn \\
&=&  \Delta h_{\mu\nu} - 4 h_{\mu\nu} -2\Gamma^\rho_{\a\mu}g^{\a\b}\nabla_\b h _{\rho\nu} - 2\Gamma^\rho_{\a\nu}g^{\a\b}\nabla_\b h _{\rho\mu} \nn \\
& & \qquad \quad \ \  -g^{\a\b}\Gamma^\sigma_{\b\mu}\Gamma^\rho_{\a\sigma}h_{\rho\nu} -g^{\a\b}\Gamma^\sigma_{\b\nu}\Gamma^\rho_{\a\sigma}h_{\rho\mu} -2g^{\a\b}\Gamma^\sigma_{\a\mu}\Gamma^\rho_{\b\nu}h_{\sigma\rho} \label{2-final}.
\end{eqnarray}
where
\begin{equation}
 \Delta h_{\mu\nu}  = \frac{1}{\sg}\partial_\alpha(g^{\alpha\beta}\partial_\beta h_{\mu\nu}) .
\end{equation}
 To obtain the 
last line in (\ref{lapspin2})  we have used (\ref{gdg}) and
rewritten the ordinary derivatives in terms of  covariant ones along with terms involving the 
Christoffel symbols. 
The next step is to look at  the $(++)$, $(--)$ and $(+-)$ components of the above equation
explicitly and repeatedly use the equations (\ref{g1}) to (\ref{g4}) and (\ref{tr-g}) to reduce the 
terms involving the Christoffel symbols in (\ref{lapspin2}) to constants which will become the 
mixing `mass matrix'. 

\vspace{.5cm}
\noindent
{\bf $(++)$\, component}
\vspace{.5cm}

Examining the $(++)$ component of (\ref{lapspin2}), we obtain
\begin{align}
\label{plusp}
\nabla ^2 h _{++} &= \Delta h_{++} - 4h_{++} \nn \\
& \quad  -4\Gamma^\xi_{++}g^{++}\nabla_+ h_{\xi+} - 4\Gamma^+_{\xi +}g^{\xi\xi}\nabla_\xi h_{++} \nn \\
& \quad  -2g^{++}\Gamma^\xi_{++}\Gamma^+_{+\xi}h_{++} - 2g^{\xi\xi}\Gamma^+_{+\xi}\Gamma^+_{+\xi}h_{++} \nn \\
& \quad  -2g^{++}\Gamma^\xi_{++}\Gamma^\xi_{++}h_{\xi\xi} - 2g^{\xi\xi}\Gamma^+_{+\xi}\Gamma^+_{+\xi}h_{++} .
\end{align}
The terms on the third line vanish because of the identity (\ref{c1}), therefore we obtain
\begin{align}
\label{pp}
\nabla ^2 h _{++} &= \Delta h_{++} - 4h_{++} + 4 \coth\xi \ (\nabla _+ h_{\xi+} - \nabla _\xi h_{++}) + 2 \cosh ^2 \xi \ h_{\xi\xi} - 2\coth ^2 \xi \ h_{++} \nn \\
&=  \Delta h_{++} - 4h_{++} - 4m h_{-+} - 2h_{--}.
\end{align}
Here we have substituted the expressions for the Christoffel symbols and the metric components
from (\ref{diamet}) and (\ref{chris}) respectively. 
To obtain the  last equality  in (\ref{plusp})  we have used 
(\ref{g4}) and the tracelessness condition (\ref{tr-g}). 

\vspace{.5cm}
\noindent
{\bf $(--)$ component}
\vspace{.4cm}
\begin{align}
\label{mm}
\nabla ^2 h _{--} &= \Delta h_{--} - 4h_{--} \nn \\
& \quad  -4\Gamma^\xi_{--}g^{--}\nabla_+ h_{\xi-} - 4\Gamma^-_{\xi-}g^{\xi\xi}\nabla_\xi h_{--} \nn \\
& \quad  -2g^{--}\Gamma^\xi_{--}\Gamma^+_{-\xi}h_{--} - 2g^{\xi\xi}\Gamma^+_{-\xi}\Gamma^+_{-\xi}h_{--} \nn \\
& \quad  -2g^{--}\Gamma^\xi_{--}\Gamma^z_{--}h_{\xi\xi} - 2g^{\xi\xi}\Gamma^+_{-\xi}\Gamma^+_{-\xi}h_{--} 
\end{align}
The manipulations here are similar to the 
$(++)$ case. The  terms on the third line vanish because of the identity (\ref{c1})
\begin{align}
\nabla ^2 h _{--} &= \Delta h_{--} - 4h_{--} + 4 \tanh \xi \ (\nabla _- h_{\xi-} - \nabla _\xi h_{--}) - 2 \sinh ^2 \xi \ h_{\xi\xi} - 2\tanh ^2 \xi \ h_{--} \nn \\
&=  \Delta h_{--} - 4h_{--} - 4m h_{+-} - 2h_{++} .
\end{align}
To obtain the last line we have used  (\ref{g3}) and the tracelessness condition (\ref{tr-g}). 

\vspace{.5cm}
\noindent
{\bf $(+-)$ component}
\vspace{.4cm}
\begin{align}
\nabla ^2 h _{+-} &= \Delta h_{+-} - 4h_{+-} \nn \\
& \quad  -2\Gamma^\xi_{++}g^{++}\nabla_+ h_{\xi-} - 2\Gamma^+_{z+}g^{\xi\xi}\nabla_\xi h_{+-} \nn \\
& \quad  -2\Gamma^\xi_{--}g^{--}\nabla_+ h_{\xi+} - 2\Gamma^-_{z-}g^{\xi\xi}\nabla_\xi h_{+-} \nn \\
& \quad  -g^{++}\Gamma^\xi_{++}\Gamma^+_{+\xi}h_{+-} - g^{\xi\xi}\Gamma^+_{\xi+}\Gamma^+_{\xi+}h_{+-} \nn \\
& \quad  -g^{--}\Gamma^\xi_{--}\Gamma^-_{-\xi}h_{-+} - 2g^{\xi\xi}\Gamma^-_{\xi-}\Gamma^-_{\xi-}h_{-+} \nn \\
& \quad  -2g^{\xi\xi}\Gamma^+_{\xi+}\Gamma^-_{\xi-}h_{-+} .
\end{align}
The terms on the fourth and fifth lines vanish because of the identity (\ref{c1})
\begin{align}
\label{pm}
\nabla ^2 h _{+-} &= \Delta h_{+-} - 4h_{+-}  \nn \\
& \quad   + 2 \coth\xi \ (\nabla _+ h_{\xi-} - \nabla _\xi h_{+-})+ 2 \tanh \xi \ (\nabla _- h_{\xi+} - \nabla _\xi h_{-+}) -2h_{+-} \nn \\
&= \Delta h_{+-} - 6h_{+-}  - 2m (h_{++}+h_{--}) .
\end{align}
We have used equations    (\ref{g1}), (\ref{g2})  to obtain the last equality. 

To summarize, the results of (\ref{pp}), (\ref{mm}) and (\ref{pm}) are the following equations. 
\begin{align}
\nabla ^2 h _{++}&=  \Delta h_{++} - 4h_{++} - 4m h_{-+} - 2h_{--} ,\\
\nabla ^2 h _{--}&=  \Delta h_{--} - 4h_{--} - 4m h_{+-} - 2h_{++} ,\\
\nabla ^2 h _{+-}&= \Delta h_{+-} - 6h_{+-}  - 2m (h_{++}+h_{--}). 
\end{align}
Note that we have reduced the action of the spin 2 Laplacian on these components 
to a scalar Laplacian together with a `mass matrix'. 
The fact that the spin $2$ Laplacian in the BTZ background
 reduces to the scalar Laplacian was noticed earlier in \cite{Sachs:2008gt,Chen:2010ik}. 
It will prove useful  to  cast these equations in the following matrix form
\begin{equation}\label{2-laplacian-matrix1}
\nabla ^2 \begin{pmatrix}
h_{++} \\
h_{+-} \\
h_{--}
\end{pmatrix} =  \Delta \begin{pmatrix}
h_{++} \\
h_{+-} \\
h_{--}
\end{pmatrix} + \begin{pmatrix}
-4 &-4m &-2 \\
-2m &-6 &-2m \\
-2 &-4m &-4
\end{pmatrix} \begin{pmatrix}
h_{++} \\
h_{+-} \\
h_{--}
\end{pmatrix}.
\end{equation}

\subsection{Solutions of the spin-2 components}
We now need to decouple these equations. In other words, diagonalize the `mass matrix' appearing in (\ref{2-laplacian-matrix1}). The linear combination which decouples the equation for the spin-1 case is \(A_+ \pm A_-\). 
This is the same linear combination which results when 
 we make the coordinate transformation
\begin{equation}
\label{cotrans}
x_1=x_+ + x_- \ , \quad x_2=x_+ - x_- \nn .
\end{equation}
On performing this linear coordinate transformation 
on the metric components $h_{++}$, $h_{--}$ and $h_{+-}$, we obtain the following transformation
 matrix
\begin{align}\label{2-coe-matrix}
\begin{pmatrix}
h_{++} \\
h_{--} \\
h_{+-}
\end{pmatrix} = \begin{pmatrix}
1 &2 &1 \\
1 &0 &-1 \\
1 &-2 &1 
\end{pmatrix}\begin{pmatrix}
h_{11} \\
h_{12} \\
h_{22}
\end{pmatrix}.
\end{align}
Note that the co-ordinate transformation  in (\ref{cotrans}) is used
just as a mathematical device to construct the linear combinations given in (\ref{2-coe-matrix}). 
It is important to emphasize that we do not 
 replace the co-ordinates in the Laplacian or in the functional dependence 
of the metric components by the definition in (\ref{cotrans}). 
Substituting the linear combinations of (\ref{2-coe-matrix}) in 
 (\ref{2-laplacian-matrix1}) results in the following decoupled equation
\begin{equation}\label{2-laplacian-matrix}
\nabla ^2 \begin{pmatrix}
h_{11} \\
h_{12} \\
h_{22}
\end{pmatrix} =\begin{pmatrix}
\Delta -4m-6 &0 &0 \\
0 &\Delta -2 &0 \\
0 &0  &\Delta +4m-6
\end{pmatrix} \begin{pmatrix}
h_{11} \\
h_{12} \\
h_{22}
\end{pmatrix}.
\end{equation}
Thus we  have diagonalized the action of the Laplacian by considering the linear 
combinations $h_{11}$, $ h_{12}$ and  $h_{22}$. 
The second order 
equations of motion of the spin 2 field which can be read out from (\ref{2ndOrder}) is given by
\begin{equation}
(\nabla^2 - m^2 + 3) h _{\mu\nu} = 0 .
\end{equation}
Thus the metric components $h_{11}, \ h_{12}$ and $h_{22}$ obey the following equations
\def\p{\partial}
\begin{align}
\label{12eq}
(\Delta -(m+2)^2 +1)h_{11} &= 0, \\
(\Delta -m^2 + 1)h_{12} &=0, \\
(\Delta -(m-2)^2 +1)h_{22} &=0.
\end{align}
Since these are the  equations of  a massive scalar, their solutions can be easily found 
We first substitute the ansatz,
\begin{equation}
h_{ij}=e^{-i(k_+ x^+ + k_-x^-)}R_{ij}.
\end{equation}
where $i, j \in\{1, 2\}$.  Note that with the definition of $x^+$ and $x^-$ given in (\ref{xpxm}) we see that the 
frequency and the momenta of these solutions 
are related to $k_+$ and $k_-$ by the following equations
\begin{equation}
 \label{fremom}
(k_+ + k_- )( r_+ - r_-) = \omega - k, \qquad ( k_+ - k_-) ( r_+ + r_-) = \omega + k .
\end{equation}
We then use the coordinate $z=\tanh^2 \xi$ to write out the 
Laplacian for each of the equations in (\ref{12eq}). This results in
\begin{align}
z(1-z)\frac{d^2 R_{11}}{dz^2} + (1-z)\frac{dR_{11}}{dz} + \Big{[} \frac{k_+^2}{4z} -  \frac{k_-^2}{4} - \frac{(m+2)^2 -1 }{4(1-z)}\Big{]}  R_{11} &= 0, \\
z(1-z)\frac{d^2 R_{12}}{dz^2} + (1-z)\frac{dR_{12}}{dz} + \Big{[} \frac{k_+^2}{4z} -  \frac{k_-^2}{4} - \frac{m^2 -1 }{4(1-z)}\Big{]}  R_{12} &= 0, \\
z(1-z)\frac{d^2 R_{22}}{dz^2} + (1-z)\frac{dR_{22}}{dz} + \Big{[} \frac{k_+^2}{4z} -  \frac{k_-^2}{4} - \frac{(m-2)^2 -1 }{4(1-z)}\Big{]}  R_{22} &= 0.
\end{align}
The solution that obeys in-going boundary conditions at the horizon is
\begin{equation}\label{spin-2-solution}
\begin{aligned}
R_{11} (z) &=e_{11} z^\alpha (1-z)^{\beta _{11}} F(a_{11}, b_{11}, c; z), \\
R_{12} (z) &=e_{12} z^\alpha (1-z)^{\beta _{12}} F(a_{12}, b_{12}, c; z), \\
R_{22} (z) &=e_{22} z^\alpha (1-z)^{\beta _{22}} F(a_{22}, b_{22}, c; z).
\end{aligned}
\end{equation}
Here $e_{ij}$s are constants which we will call as the polarizations. 
The parameters of the functions defined in (\ref{spin-2-solution})   are given by
\begin{eqnarray}
\alpha = \frac{-ik_+}{2}, & & c = 1+ 2\alpha, \\ \nonumber
\beta _{11} = \frac{1}{2} (m+3), \qquad 
\beta _{12} = \frac{1}{2} (m+1), \qquad & &
\beta _{22} = \frac{1}{2} (m-1) , \\ \nonumber
a_{ij}  = \frac{k_+ - k_-}{2i} + \beta _{ij}, & \qquad&  
b_{ij} = \frac{k_+ + k_-}{2i} + \beta _{ij}  .
\end{eqnarray}
Note that in each of the cases one can choose an alternate form of the solution due to the following identity obeyed by the hypergeometric function
\begin{equation}
\label{hypident}
 F(a, b, c; z) = ( 1-z) ^{c-a-b} F( c-a, c-b, c;z) .
\end{equation}

\vspace{.5cm}
\noindent
{\bf Determining the polarization constants}
\vspace{.5cm}

Our next task is to fix the polarization constants or coefficients \(e_{ij}\) appearing in the solutions 
given in (\ref{spin-2-solution}). 
Since these are constants through out space time, 
their values can be found by looking at the behaviour at any specific point or surface. 
 We will see that these constants are easily determined by examining the solutions near the horizon
$z\rightarrow 1$. 
This approach is different from the one followed by 
 \cite{Das:1999pt} for the case of vectors. 
There the values of   the polarization constants  were found using the recursion properties 
the of hypergeometric functions. 
We will see that the method developed here is also easily generalized to the 
case of arbitrary spin.

The near-horizon $(z\rightarrow 0)$ behaviour of these solutions is
\begin{equation}
R_{ij}\rightarrow e_{ij} z^{\alpha} e^{-i(  k_+x^+ k_-x^-)}, \qquad \hbox{as} \qquad z \rightarrow 0.
\end{equation}
Now from the relation (\ref{2-coe-matrix}) we see that the behaviour of 
of the solutions $h_{++}, h_{+-}, h_{--}$ near the horizon is given by
\begin{equation}
h_{\hat\mu\hat \nu} \rightarrow e_{\hat\mu\hat\nu } z^\alpha  e^{-i(  k_+x^+ k_-x^-)},
 \qquad \hbox{as} \qquad z \rightarrow 0.
\end{equation}  
where, $\hat \mu, \hat \nu \in \{ +, -\}$ and 
\begin{align}\label{2-coe-matrixc}
\begin{pmatrix}
e_{++} \\
e_{--} \\
e_{+-}
\end{pmatrix} = \begin{pmatrix}
1 &2 &1 \\
1 &0 &-1 \\
1 &-2 &1 
\end{pmatrix}\begin{pmatrix}
e_{11} \\
e_{12} \\
e_{22}
\end{pmatrix}.
\end{align}
We will now determine the relations between the coefficients $e_{\hat\mu\hat\nu}$ by 
examining the equation (\ref{CS-2}) near the horizon. 
The $(\xi +)$ and $( \xi - )$ components of equation (\ref{CS-2}) gives rise to the 
following relations
\begin{align}
-m\frac{\sqrt{z}}{1-z} h_{\xi+} &= \p _- h_{++} -\p _+ h_{+-} + \frac{\sqrt{z}}{1-z}h_{\xi -}, \\
-m\frac{\sqrt{z}}{1-z} h_{\xi-} &= \p _- h_{+-} -\p _+ h_{--} + \frac{\sqrt{z}}{1-z}h_{\xi +}.
\end{align}
Near the horizon the above equations can be used to determine the form 
of $h_{\xi+}$ and $h_{\xi-}$ in terms of $e_{\hat \mu\hat \nu}$. This can be 
written in terms of the following matrix equation
\begin{equation}\label{2-matrix}
\begin{pmatrix}
m & 1 \\
1  & m 
\end{pmatrix}
\begin{pmatrix}
 h_{\xi+} \\
 h_{\xi-}
\end{pmatrix}= i\begin{pmatrix}
k_-e_{++}-k_+e_{+-} \\
k_-e_{+-}-k_+e_{--}
\end{pmatrix}z^{\alpha- \frac{1}{2} } e^{-i(  k_+x^+ k_-x^-)}.
\end{equation}
Thus $h_{\xi\pm}$ has the following  behaviour
\begin{equation}
 h_{\xi \hat \mu } \rightarrow  e_{\xi \hat \mu} z ^{\alpha - \frac{1}{2} }   e^{-i(  k_+x^+ k_-x^-)},
\quad 
\hbox{as}\quad  z \rightarrow 0.
\end{equation}
Let us now examine 
the equation (\ref{g2}) in detail
\begin{align}
-mh_{--} &= \frac{-1}{\sqrt{z}} (\p_\xi h_{+-} - \p_-h_{\xi -} ) +h_{+-} \nn \\
   &= - \frac{1}{\sqrt{z}} (2\sqrt{z}(1-z)\p _z h_{+-} -\p_+ h_{\xi -}) + h_{+-} \nn.
\end{align}
Near the horizon, the first two terms on the right hand side have the behaviour $\sim z^{\alpha-1}$ while other terms have the behaviour $\sim z^\alpha$. 
Thus to the leading order as $z\rightarrow0$, we must have the following relation
\begin{equation}
\label{epm}
e_{+-} =e_{\xi-}.
\end{equation}
Similarly if we investigate the near-horizon behaviour of equation (\ref{g4}), we obtain the relation
\begin{equation}
\label{epp}
e_{++} =e_{\xi+}.
\end{equation}
Substituting these (\ref{epm}) and (\ref{epp})  in (\ref{2-matrix}) we obtain the 
following homogeneous equation relating the three coefficients $e_{\hat\mu\hat\nu}$,
\begin{equation}
\begin{pmatrix}
m-ik_-    &1+ik_+    &0   \\
1            &m-ik_-     &ik_+ 
\end{pmatrix}\begin{pmatrix}
e_{++} \\
e_{+-} \\
e_{--}
\end{pmatrix}=0.
\end{equation}
We can write this equation in 
in terms $e_{11},e_{12}$ and $e_{22}$ using (\ref{2-coe-matrix}), this results in
\begin{equation}
\begin{pmatrix}
m-ik_-    &1+ik_+    &0   \\
1            &m-ik_-     &ik_+ 
\end{pmatrix}\begin{pmatrix}
1 &2 &1 \\
1 &0 &-1 \\
1 &-2 &1 
\end{pmatrix}\begin{pmatrix}
e_{11} \\
e_{12} \\
e_{22}
\end{pmatrix} = 0.
\end{equation}
The easiest way to write one of the polarization constants in terms of the other is to 
multiply the above equation 
by the  spin-1 coefficient transformation matrix. Thus we have 
\begin{eqnarray}
& & 
\begin{pmatrix}
1 &1 \\
1 &-1 \\
\end{pmatrix} \begin{pmatrix}
m-ik_-    &1+ik_+    &0   \\
1            &m-ik_-     &ik_+ 
\end{pmatrix}\begin{pmatrix}
1 &2 &1 \\
1 &0 &-1 \\
1 &-2 &1 
\end{pmatrix}\begin{pmatrix}
e_{11} \\
e_{12} \\
e_{22}
\end{pmatrix} =  \\ \nonumber
& & \begin{pmatrix}
m+1+i(k_+ - k_-)  &\ m+1-i(k_- +k_+)   &\ 0  \\
0                           &\ m-1+i(k_+ -k_-)     &\ m-1-i(k_+ +k_-)  
\end{pmatrix}  \begin{pmatrix}
e_{11} \\
e_{12} \\
e_{22}
\end{pmatrix} = 0 .
\end{eqnarray}
These result in the following recursion relations
\begin{align}
(m+1+i(k_+ - k_-))e_{11}=-(m+1-i(k_- +k_+))e_{12} ,\\
( m-1+i(k_+ -k_-) )e_{12} = -( m-1-i(k_+ +k_-) )e_{22} .
\end{align}
Now we can easily obtain the polarization constants $e_{11}, e_{12}$ in terms of 
$e_{22}$ easily.  We therefore obtain the following relations
\begin{align}
e_{12} &= -\frac{( m+1-i(k_+ +k_-) )}{( m+1+i(k_+ -k_-) )} e_{22}, \\
e_{11}&=-\frac{(m-1-i(k_+ +k_-))}{(m-1+i(k_+ - k_-))}e_{12}, \nn \\
 &=\frac{( m-1-i(k_+ +k_-) )(m+1-i(k_+ +k_-))}{( m-1+i(k_+ -k_-) )(m+1+i(k_+ - k_-))} e_{22}  .
\end{align}

\subsection{Quasinormal modes}

Now we have all the ingredients necessary to obtain the quasinormal modes of the spin 2
field in the BTZ background. To do this 
we examine the behaviour of the solutions (\ref{spin-2-solution}) at the boundary   $z\rightarrow1$. 
Let us consider first the case of the solution $R_{11}$. Near the boundary the dominant 
behaviour  for the case $m>0$ is given by 
\begin{align}
\label{r11bc}
R_{11} (z) &\simeq e_{11} z^\alpha (1-z)^{\frac{-m-1}{2}} \frac{\Gamma(c)\Gamma(a_{11} +b_{11} - c)}{\Gamma(a_{11})\Gamma(b_{11})} \nn \\
&= \frac{4 e_{11} z^\alpha (1-z)^{\frac{-m-1}{2}} \Gamma(c)\Gamma(a_{11} +b_{11} - c)}{( m-1-i(k_+ +k_-) )(m+1-i(k_+ +k_-))\Gamma(a_{11})\Gamma(b_{11}-2)} \nn \\
&= \frac{4e_{22} z^\alpha (1-z)^{\frac{-m-1}{2}}\Gamma(c)\Gamma(a_{11} +b_{11} - c)}{( m-1+i(k_+ -k_-) )(m+1+i(k_+ - k_-))\Gamma(a_{11})\Gamma(b_{11}-2)} .
\end{align}
Quasinormal modes are obtained by imposing vanishing 
Dirichlet conditions at the boundary. This implies 
we look for the condition in which $R_{11}$ near the boundary vanishes.  
The zeros  of the function in (\ref{r11bc})  occur at the locations, $a_{11}=-n$ and $b_{11}-2=-n$ 
with $n = 0, 1, 2, \cdots$, which in terms of 
the momenta are given by
\begin{equation}
i(k_+ + k_-) = 2n+\hat\Delta -2,  \qquad i(k_+ - k_-) = 2n+\hat\Delta +2, \qquad n = 0, 1, 2, \cdots .
\end{equation}
Here, we have used the relation
\begin{equation}
 \hat \Delta=1+m. \nn
\end{equation}
which results from (\ref{conf-dim}), note that we have assumed $m>0$. 
Now the behaviour of the function $R_{12}$ near the boundary is given by
\begin{align}
\label{r12bc}
R_{12} (z) &\simeq e_{12} z^\alpha (1-z)^{\frac{-m+1}{2}} \frac{\Gamma(c)\Gamma(a_{12} +b_{12} - c)}{\Gamma(a_{11})\Gamma(b_{11})} \\
&=\frac{ 2e_{12} z^\alpha (1-z)^{\frac{-m+1}{2}} \Gamma(c)\Gamma(a_{12} +b_{12} - c)}{( m+1-i(k_+ +k_-) )\Gamma(a_{11})\Gamma(b_{11}-1)} \\
&=-\frac{ 2e_{22} z^\alpha (1-z)^{\frac{-m+1}{2}} \Gamma(c)\Gamma(a_{12} +b_{12} - c)}{( m+1+i(k_+ -k_-) )\Gamma(a_{11})\Gamma(b_{11}-1)} .
\end{align}
The zeros of $R_{12}$ therefore occur at the locations, $a_{12}=-n$ and $b_{12}-1=-n$. 
In terms of the momenta
they are at
\begin{equation}
i(k_+ + k_-) = 2n+\hat\Delta -2 \ , \qquad i(k_+ - k_-) = 2n+\hat \Delta, \qquad n =0, 1, 2 , \ldots .
\end{equation}
Finally the component $R_{22}$ near the boundary is given by
\begin{align}
\label{r22bc}
R_{22} (z) &\simeq e_{22} z^\alpha (1-z)^{\frac{-m+3}{2}} \frac{\Gamma(c)\Gamma(a_{22} +b_{22} - c)}{\Gamma(a_{22})\Gamma(b_{22})}.
\end{align}
The zeros of $R_{22}$ thus occur at the locations, $a_{22}=-n$ and $b_{22}=-n$, which in terms of momenta are given by
\begin{equation}
i(k_+ + k_-) = 2n+\hat\Delta -2 \ , \qquad i(k_+ - k_-) = 2n+\hat\Delta -2, \quad n= 0, 1, 2, \ldots.
\end{equation}
Vanishing Dirichlet conditions  at the boundary for  all  the components of spin-2 field 
require all the  functions in (\ref{r11bc}), (\ref{r12bc}) and (\ref{r22bc}) to vanish 
simultaneously  as $z\rightarrow 1$. The quasinormal modes are thus given by the common set of zeros of all the components. These are
\begin{equation}
i(k_+ + k_-) = 2n+\hat\Delta -2 \ , \qquad i(k_+ - k_-) = 2n+\hat \Delta +2, \qquad n =0, 1, 2, \ldots  .
\end{equation}
Writing the  above relations in terms of frequency, momenta and the temperatures using 
(\ref{fremom}), we obtain the following set of quasinormal modes
\begin{equation}
\begin{aligned}
\omega _L &=  k+2\pi T_L(k_+ +k_-) \\
&=  k-2\pi iT_L(2n + \hat \Delta -2),
\end{aligned}\qquad
\begin{aligned}
\omega _R &= -k+2\pi T_R(k_+ -k_-) \\ 
&= -k-2\pi iT_R (2n+\hat \Delta +2).
\end{aligned}
\end{equation}
with $n =0, 1, 2, \ldots$. 
Comparing this to equation 
(\ref{polesl}) we see that the quasinormal modes for the spin 2 field precisely coincide with the 
poles of the corresponding retarded two point function.  Notice that we have 
obtained $h_R -h_L =+2$. The case $h_R -h_L =-2$ arises when we consider the 
situation with $m<0$. 

\def\p{\phi}
\def\m{\mu}
\def\n{\nu}
\def\G{g^{\a\b}}

\def\cd{\cdots}
\def\MU{\m_3\cdots\m_s}
\def\MP{\m_3\cdots\m_p}
\def\NQ{\nu_3\cdots\n_q}
\def\U{\m_1\cd\m_p}
\def\V{\nu_1\cd\nu_q}

\section{The spin $s$ case}

In this section we will generalize the same procedure which we developed for the 
spin 2 case for arbitrary spins. 
The first order equation which we will repeatedly use for the reduction of the 
spin $s$ Laplacian to the scalar Laplacian is the following
\begin{equation}\label{1stOrderForS}
\epsilon_{\m_1} ^{\ \alpha \beta} \nabla _\alpha \Phi_{\beta \m_2\m_3\cdots\m_s} = -m \phi_{\m_1 \m_2\m_3\cdots\m_s}.
\end{equation}
The $(\pm\m_2\MU)$ components of the above equation can be written as
\begin{align}
\label{s-CS} \pm \frac{g_{\pm \pm}}{\sg}  (\nabla_\xi \Phi_{\mp \m_2\m_3\cdots\m_s}-\nabla_\mp \Phi_{\xi \m_2\m_3\cdots\m_s}) =  -m  \phi_{\pm \m_2\m_3\cdots\m_s} .
\end{align}
We will also require the 
use of the  tracelessness condition which is
\begin{align} \label{tr-s}
\Phi_{\xi\xi\MU} -\csch ^2 \xi \ \Phi_{++\MU} +\sech^2 \xi \ \Phi_{--\MU}  = 0 .
\end{align}

\subsection{Reduction of the spin-$s$ Laplacian to the scalar Laplacian}

After some straightforward manipulations and the use of (\ref{c0}),
the action of the Laplacian on an arbitrary spin field, $\Phi_{\m_1\m_2\MU}$ can be
written as
\begin{align}
\nabla ^2 \Phi _{\m_1\m_2\MU} &= \Delta  \Phi _{\m_1\m_2\MU}  \nn \\
&\quad -2 ( \hat\delta_{\mu_1}^\sigma  \Phi _{\sigma\m_2\MU}
+ \hat\delta_{\mu_2}^\sigma\Phi_{\m_1\sigma\MU} 
+ \cdots +  \hat\delta_{\mu_s}^\sigma\Phi_{\mu_1\mu_2\cdots \mu_{s-1}\sigma} ) \nn \\
&\quad -2\G (\Gamma^\eta _{\b \m_1}\nabla _\a \Phi _{\eta\m_2\MU} + \Gamma^\eta _{\b \m_2}\nabla _\a \Phi _{\m_1\eta\MU}+\cdots \nn \\
&\quad\qquad\qquad\qquad+\Gamma^\eta _{\b \m_s}\nabla _\a  \Phi _{\m_1\cdots\m_{s-1}\eta})  \nn \\
&\quad - \G \Gamma ^\sigma _{\a\eta} (\Gamma^\eta _{\b \m_1} \Phi _{\sigma\m_2\MU}+ \Gamma^\eta _{\b \m_2} \Phi _{\m_1\sigma\MU} +\cdots+\Gamma^\eta _{\b \m_s} \Phi _{\m_1\cdots\m_{s-1}\sigma})\nn \\
&\quad - 2\G \sum_{i, j, i\neq j}\Gamma^\eta _{\b \m_i} \Gamma^\sigma _{\a \m_j}  
\Phi _{\eta\sigma\cdots \check \mu_i\cdots \check \mu_j \cdots \mu_s} .
\label{S-final}
\end{align}
Here, \(\Delta\) is the scalar Laplacian or
\begin{equation}
  \Delta  \Phi _{\m_1\m_2\MU}  = \frac{1}{\sg}\partial_\alpha(g^{\alpha\beta}\partial_\beta  \Phi _{\m_1\m_2\MU}) .
  \end{equation}
 In the last line of (\ref{S-final}), the $\check\mu_i, \check\mu_j$ refer to the indices which are missing and 
 $i, j$ run over $1$ to $s$. 
Just as for the spin 2 case, to determine the quasinormal modes it is sufficient to focus on 
components which have only $+$s or $-$s. This is because, the other components
can be determined by them and can be shown by the same argument used for the 
spin 2 case to be  suppressed near the boundary. 
Therefore, let us now examine 
 the action of the Laplacian on the object, $\Phi _{\U\V}$ with $\U=+\cd+$ and $\V=-\cd-$.
  with $p+q =s$.  This is given by 
\begin{align}
&\nabla ^2 \Phi _{\U\V} = \Delta  \Phi _{\U\V} -2s \Phi _{\U\V} \nn \\
& -2\G (\Gamma^\eta _{\b \m_1}\nabla _\a \Phi _{\eta\m_2\MP\V} + \Gamma^\eta _{\b \m_2}\nabla _\a \Phi _{\m_1\eta\MP\V}+\cdots+\Gamma^\eta _{\b \m_s}\nabla _\a  \Phi _{\m_1\cdots\m_{s-1}\eta\V} \nn \\
&\quad \quad \ +\Gamma^\eta _{\b \nu_1}\nabla _\a \Phi _{\U\eta\nu_2\NQ} + \Gamma^\eta _{\b \nu_2}\nabla _\a \Phi _{\U\nu_1\eta\NQ}+\cdots+\Gamma^\eta _{\b \nu_s}\nabla _\a  \Phi _{\U\nu_1\cdots\nu_{s-1}\eta})  \nn \\
&- \G \Gamma ^\sigma _{\a\eta}  (\Gamma^\eta _{\b \m_1} \Phi _{\sigma\m_2\MP\V} + \Gamma^\eta _{\b \m_2} \Phi _{\m_1\sigma\MP\V}+\cdots+\Gamma^\eta _{\b \m_s}  \Phi _{\m_1\cdots\m_{s-1}\sigma\V} \nn \\
&\quad \quad \quad \ +\Gamma^\eta _{\b \nu_1} \Phi _{\U\sigma\nu_2\NQ} + \Gamma^\eta _{\b \nu_2} \Phi _{\U\nu_1\sigma\NQ}+\cdots+\Gamma^\eta _{\b \nu_s}  \Phi _{\U\nu_1\cdots\nu_{s-1}\sigma})  \nn \\
& - 2\G \left(
\sum_{i, j=1; i\neq j}^p  \Gamma^\eta _{\b \m_i} \Gamma^\sigma _{\a \m_j}  
\Phi _{\eta\sigma\mu_3\cdots\check\mu_i\cdots\check\mu_j\cdots\mu_p\nu_1\cdots \nu_q } \right. 
+ \sum _{i=1, \ j=1} ^{i=p, \ j=q}  \Gamma^\eta _{\b \m_i} \Gamma^\sigma _{\a \nu_j} \Phi _{\eta\sigma\m_1\cd\check\m_{i}\cd\m_{p}\nu_1\cd\check\nu_{j}\cd\nu_{q}}  \nn \\
& \qquad \qquad\qquad\qquad\qquad \left.   +\sum_{i, j=1;i\neq j}^ q\Gamma^\eta _{\b \nu_i} \Gamma^\sigma _{\a \nu_j}  \Phi _{\eta\sigma\mu_1\cdots \mu_p\nu_1\cdots \check\nu_i\cdots \check\nu_j \cdots \nu_q} \right) .
\end{align}

Let's now introduce some notation, by `$(p)$' we mean $p$ number of $+$  indices and by `$(q)$' we mean $q\ (=s-p)$ number of $-$  indices. Here's an example, with a field of \(s=5\)
\begin{equation}
\Phi _{(2)(3)} = \Phi_{++---} .\nn
\end{equation}
For the case  $\m_1,\m_2,\cd,\m_p=+$ and $\nu_1,\nu_2,\cd,\nu_q=-$ and using  the fact that $\Phi _{\U\V}$ is a completely symmetric tensor we obtain
\begin{align}
\nabla ^2 \Phi _{(p)(q)} =& \ \Delta  \Phi _{(p)(q)} -2s \Phi _{(p)(q)} \nn \\
& -2\G (p\Gamma^\eta _{\b +}\nabla _\a \Phi _{\eta(p-1)(q)} + q\Gamma^\eta _{\b -}\nabla _\a \Phi _{\eta(p)(q-1)})\nn \\
&  - \G \Gamma ^\sigma _{\a\eta}  (p\Gamma^\eta _{\b +} \Phi _{\sigma(p-1)(q)} + q\Gamma^\eta _{\b -} \Phi _{\sigma(p)(q-1)}) \nn \\
&  - 2\G \Big{(} \frac{p(p-1)}{2} \Gamma^\eta _{\b +} \Gamma^\sigma _{\a +}  \Phi _{\eta\sigma(p-2)(q)} +  \frac{q(q-1)}{2} \Gamma^\eta _{\b -} \Gamma^\sigma _{\a -}  \Phi _{\eta\sigma(p)(q-2)} \nn \\
& \qquad \qquad +  pq \Gamma^\eta _{\b +} \Gamma^\sigma _{\a -} \Phi _{\eta\sigma(p-1)(q-1)} \Big{)}  .
\end{align}
Now writing out the terms that contribute we obtain
\begin{align}
\nabla ^2 \Phi _{(p)(q)} =& \ \Delta  \Phi _{(p)(q)} -2s \Phi _{(p)(q)} \nn \\
& -2p(g^{++}\Gamma^\xi _{+ +}\nabla _+ \Phi _{\xi(p-1)(q)} +g^{\xi\xi}\Gamma^+ _{\xi+}\nabla _\xi \Phi _{+(p-1)(q)} ) \nn \\
& -2q(g^{--}\Gamma^\xi _{--}\nabla _- \Phi _{\xi(p)(q-1)} +g^{\xi\xi}\Gamma^- _{\xi-}\nabla _\xi \Phi _{-(p)(q-1)} ) \nn \\
& - p (g^{++}\Gamma ^+ _{+\xi} \Gamma^\xi _{++} \Phi _{+(p-1)(q)}+g^{\xi\xi}
\Gamma ^+ _{\xi+} \Gamma^+ _{\xi+} \Phi _{+(p-1)(q)}) \nn \\
& - q (g^{--}\Gamma ^- _{-\xi} \Gamma^\xi _{--} \Phi _{-(p)(q-1)}+g^{\xi\xi}\Gamma ^- _{\xi-} \Gamma^- _{\xi-} \Phi _{-(p)(q-1)})\nn \\
& -  p(p-1) (g^{++} \Gamma^\xi _{+ +} \Gamma^\xi _{+ +}  \Phi _{\xi\xi(p-2)(q)} + g^{\xi\xi} \Gamma^+ _{\xi +} \Gamma^+ _{\xi +} \Phi _{++(p-2)(q)}) \nn \\
& -  q(q-1) (g^{--} \Gamma^\xi _{--} \Gamma^\xi _{--}  \Phi _{\xi\xi(p)(q-2)} + g^{\xi\xi} \Gamma^- _{\xi -} \Gamma^- _{\xi -} \Phi _{--(p)(q-2)}) \nn \\
& - 2pq \ g^{\xi\xi} \Gamma^+ _{\xi +} \Gamma^- _{\xi-} \Phi _{+-(p-1)(q-1)} .
\end{align}
The terms in the fourth and fifth lines vanish because of (\ref{c1}). We now substitute the 
values of the Christoffel symbols and the relevant metric components to obtain
\begin{align}
\nabla ^2 \Phi _{(p)(q)}&= \Delta \Phi _{(p)(q)} - 2s\Phi _{(p)(q)}  \nn \\
& \ \ +2p \coth\xi (\nabla _+ \Phi _{\xi(p-1)(q)} - \nabla _\xi \Phi _{+(p-1)(q)} ) \nn \\
& \ \ +2q \tanh\xi (\nabla _- \Phi _{\xi(p)(q-1)} - \nabla _\xi \Phi _{-(p)(q-1)} ) \nn \\
& \ \ -  p(p-1) (-\cosh ^2\xi \ \Phi _{\xi\xi(p-2)(q)} +  \coth ^2 \xi \ \Phi _{++(p-2)(q)}) \nn \\
& \ \ -  q(q-1) (\sinh ^2\xi \ \Phi _{\xi\xi(p)(q-2)} + \tanh ^2 \xi \ \Phi _{--(p)(q-2)}) \nn \\
& \ \ - 2pq  \Phi _{+-(p-1)(q-1)} .
\end{align}
Using the first order equation (\ref{s-CS}) and the tracelessness condition (\ref{tr-s}), we get 
\begin{align}
\nabla ^2 \Phi _{(p)(q)} &= \Delta  \Phi _{(p)(q)} -2s \Phi _{(p)(q)} - 2pm \Phi_{(p-1)(q+1)} -2qm \Phi_{(p+1)(q-1)} \nn \\
& \ \  \  -p(p-1)\Phi_{(p-2)(q+2)} -q(q-1) \Phi_{(p+2)(q-2)} - 2 pq \Phi _{(p)(q)} .
\end{align}
Since $p+q=s$, the equation can also be written as
\begin{align}\label{laplacian-on-s}
\nabla ^2 \Phi _{(p)(s-p)} = \Delta  \Phi _{(p)(s-p)} -2s \Phi _{(p)(s-p)} - 2pm \Phi_{(p-1)(s-p+1)} -2(s-p)m \Phi_{(p+1)(s-p-1)} \nn \\
\ \  -p(p-1)\Phi_{(p-2)(s-p+2)} -(s-p)(s-p-1) \Phi_{(p+2)(s-p-2)} - 2 p(s-p) \Phi _{(p)(s-p)} .
\end{align}
Note that we have now obtained a set of closed equations for the components of 
the metric with only the boundary indices $+$ or $-$. 

\subsection{Solutions of the spin-$s$ components}

In this section we will solve the set of 
 $(s+1)$ coupled equations which we have derived in  (\ref{laplacian-on-s}). 
From now on we will suppress the second label `$(q)$' for the number of $-$s as it is understood we are working with 
a rank $s$ symmetric tensor. 
The  set of equations in (\ref{laplacian-on-s})  can also be written in the form of a matrix as,
\begin{equation}\label{lap-scalar&mass}
\nabla ^2 \Phi _{(p)} = \Delta  \Phi _{(p)} + M^{(s)}_{pr}\Phi_{(r)}.
\end{equation}
Here, the `mass matrix' $M^{(s)}_{pq}$ is defined as
\begin{align}
M^{(s)}_{pr}=&-(2 s + 2 p (s - p)) \delta _{p,r} -2mp\delta_{p-1,r} -2m(s-p)\delta _{p+1,r} \nn \\
& -p(p-1)\delta_{p-2,r} -(s-p)(s-p-1)\delta_{p+2,r} .
\end{align}
Written out in the explicit matrix form, \(M^{(s)} \) looks like
\begin{equation}\label{mass-matrixform}
M^{(s)}={\footnotesize \begin{pmatrix}
-2s 	&-2sm 	&-s(s-1) 	 &0                 &0 &\cdots &\cdots &\cdots\\
-2m  &-(4s-2)  &-2m(s-1) &-(s-1)(s-2) &0 &\cdots &\cdots &\cdots\\
-2		 &-4m      &-(6s-8)    &-2m(s-2)    &-(s-2)(s-3) &\cdots &\cdots &\cdots\\
0       &-6         &-6m         &-(8s-18)     &-2m(s-3) &\cdots &\cdots &\cdots\\
0       &0           &-8            &-8m           &-(10s-32)  &\cdots &\cdots &\cdots\\
\hdotsfor[2]{8}\\
\hdotsfor[2]{8}\\
\cdots &\cdots &\cdots &\cdots &\cdots &\cdots &-(4s-2) &-2m\\
\cdots &\cdots &\cdots &\cdots &\cdots &\cdots &-2sm &-2s 
\end{pmatrix} } .
\end{equation}
Note that this is a $(s+1) \times (s+1) $ matrix. 

\subsubsection*{Diagonalization of the mass matrix}

From our study of the $s=2$ case we have seen that we could 
diagonalize $M^{(2)}_{pq}$ by considering the linear combinations of the 
components  obtained   using the coordinate transformation
\begin{equation}
x_1=x_+ + x_- \ , \quad x_2=x_+ - x_- \nn .
\end{equation}
We will see that we can use the same  transformation to diagonalize the `mass matrix' but  generalized 
for the arbitrary spin case. 
This motivates us to consider the following linear combination
\begin{equation}
\hat \Phi_{[p]}=\sum_{a=0}^{p}\sum_{b=0}^{s-p} (-1)^b \binom{p}{a} \binom{s-p}{b} \Phi_{(s-a-b)}.
\end{equation}
where $\hat \Phi$ is a symmetric rank $s$ tensor with indices $1$ or $2$. 
By `$[p]$' we mean, there are $p$ number of $1$s in the component of $\hat \Phi$ we are considering. This automatically implies that the are $s-p$ number of $2$s in the components. We will show that this transformation diagonalizes the matrix $M^{(s)}_{pq}$. Let us write $\Phi_{[p]}$ as 
\begin{equation}\label{transformation}
\hat \Phi_{[p]}=\sum _{q=0}^{s} T^{(s)}_{pq} \Phi_{(q)} .
\end{equation}
We can then write a generating function for the coefficient $T^{(s)}_{pq}$ by proceeding 
as follows. Consider the polynomial
\begin{eqnarray}
 P_{(s, p)} (x)  &=& 
\sum_{q=0}^{s}T^{(s)}_{pq} x^q = \sum_{a=0}^{p}\sum_{b=0}^{s-p} (-1)^b \binom{p}{a} \binom{s-p}{b} x^{(s-a-b)}, \\ \nonumber
&=& (x+1)^{p} (x-1)^{s-p} .
\end{eqnarray}
Thus  $T^{(s)}_{pq}$ is the coefficient of $x^q$ of the function above. A formal expression for $T^{(s)}_{pq}$ can then be obtained by a Taylor series expansion and be expressed as a contour integral.
\begin{equation}\label{T-expression}
T^{(s)}_{pq}=\frac{1}{2\pi i} \oint \frac{dx}{x^{q+1}} P_{(s, p)} (x) .
\end{equation} 
It can be  shown that this transformation matrix $T^{(s)}$ obeys the following identities. 

\vspace{.5cm}
\noindent
{\bf Identity 1:}
\vspace{.5cm}
\begin{equation}\label{claim-0}
\sum_{q=0}^{s} T^{(s)}_{pq}T^{(s)}_{qr} = 2^s \delta_{qr} .
\end{equation}

\vspace{.5cm}
\noindent
{\bf Identity 2:} 
\begin{equation}\label{claim-1}
\left( T^{(s)}M^{(s)}[T^{(s)}]^{-1} \right)_{pq}=-((2p-s)^2+2m(2p-s)+s)\delta_{pq} ,
\end{equation}
i.e., $T^{(s)}M^{(s)}[T^{(s)}]^{-1}$ is diagonal. 
\vspace{.5cm}

\noindent
The proofs of these identities are provided in Appendix C.

Using the transformation $T^{(s)}$,  the Laplacian acting on  $\hat \Phi _{[p]}$ reduces to
\begin{align}\label{decoupled-s}
\nabla^2\hat \Phi _{[p]} &= (\Delta - (2p-s)^2-2m(2p-s)-s)\hat \Phi _{[p]} .
\end{align}
Now including the spin dependent mass shift in the second order equations of motion given in
(\ref{2ndOrder}) we obtain the following equations of motion for  $\hat \Phi _{[p]}$.
\def\P{{[p]}}
\begin{align}\label{hypgeo-s-1}
(\Delta - (m+2p-s)^2+1) \hat \Phi _{[p]}  &=0  .
\end{align}
We make the ansatz,
\begin{equation}
 \Phi _{[p]} =e^{- i(k_+x^+ + k_-x^-)}R_{[p]} (\xi).
\end{equation}
Substituting this ansatz and writing out the scalar Laplacian we obtain
\begin{align}\label{hypgeo-s-2}
z(1-z)\frac{d^2 R_{[p]} }{dz^2} + (1-z)\frac{dR_{[p]} }{dz} + \Big{[} \frac{k_+^2}{4z} -  \frac{k_-^2}{4} - \frac{(m+2p-s)^2 -1 }{4(1-z)}\Big{]}  R_{[p]} &= 0 .
\end{align}
where $z=\tanh^2 \xi$. 
The solution that obeys the ingoing boundary conditions at the horizon is given by 
\begin{align}\label{S-soln}
R_{[p]}  (z) &=e_{[p]}  z^\alpha (1-z)^{\beta _{[p]} } F(a_{[p]} , b_{[p]} , c; z) .
\end{align}
where
\begin{eqnarray}\label{aabeecees}
\alpha &= & \frac{-ik_+}{2}, \qquad \quad c = 1+ 2\alpha \\ \nonumber
\beta _{[p]}  &=&  \frac{1}{2} (1+ m +2p-s) \\ \nonumber
a_{[p]}  &=& \frac{k_+ - k_-}{2i} + \beta _{[p]}, \qquad 
b_{[p]}  = \frac{k_+ + k_-}{2i} + \beta _{[p]}  .
\end{eqnarray}
As mentioned for the spin 2 case one can also work with the other choice of hypergeometric function
given by the relation (\ref{hypident}) which leads to equivalent results.

\subsubsection*{Determining  the polarization constants $e_{[p]}$}

We shall now find the coefficients or polarization constants, $e_{[p]}$ along the same 
lines of what we had done for the $s=2$ case. 
Since these are constants we can determine them at any point in space.
As we have done in the earlier section, we will see that it will be easy to determine
them by examining the behaviour of the solutions near the horizon. 
 
Near the horizon  $z\rightarrow0$, the behaviour of the solutions can be 
obtained by examining (\ref{S-soln}). Therefore we have
\begin{equation}\label{horizon-01}
\hat\Phi_\P (z)\rightarrow  e_\P z^\alpha e^{i(k_+x^+ + k_-x^-)}, \qquad z\rightarrow 0.
\end{equation}
Since $\Phi_{(p)}$ is just a linear combination of $\hat\Phi_\P(z)$ we also have
\begin{equation}
\Phi_{(p)}(z)\rightarrow  e_{(p)} z^\alpha e^{i(k_+x^+ + k_-x^-)}, \qquad z\rightarrow 0.
\end{equation}
where
\begin{equation}
\label{poltrans}
e_{(p)} = \frac{1}{2^s}\sum_{q=0}^sT_{pq} ^{(s)}e_{[q]}.
\end{equation}
Here we have used (\ref{claim-0}) to write out the inverse transformation. 
Let us now examine the `${\xi(p)(q)}$'-component\footnote{It's implied that  $q=s-p-1$.} of  the equation (\ref{1stOrderForS}) and
  see its consequences 
\begin{equation}
 \epsilon_{\xi}^{\ \alpha\beta} \nabla_\alpha \Phi_{\beta(p)(q)} = -m\Phi_{\xi(p)(q)} .
 \end{equation}
 Expanding this equation and after some simple manipulations we obtain
 \begin{equation}
 \partial_- \Phi_{(p+1)(q)}- \partial_+ \Phi_{(p)(q+1)} = -p\Gamma^\xi_{++}\Phi_{\xi(p-1)(q+1)}-m\sg\Phi_{\xi(p)(q)}+q\Gamma^\xi_{--}\Phi_{\xi(p+1)(q-1)}  .
\end{equation}
The near-horizon behaviour of the above equation is
\begin{align}\label{horizon-1}
-ik_- \Phi_{(p+1)(q)}+&ik_+ \Phi_{(p)(q+1)}\nn \\
 &\simeq -\sqrt{z}(p\Phi_{\xi(p-1)(q+1)}+m\Phi_{\xi(p)(q)}-q\Phi_{\xi(p+1)(q-1)}) .
\end{align}
Thus, it's evident that the near-horizon behaviour of $\Phi_{\xi(p)(q)}$ is
given by
\begin{equation}
\Phi_{\xi(p)(q)}\rightarrow  e_{\xi(p)(q)} z^{\alpha- \frac{1}{2}} e^{-i (k_+x^+ + k_-x^-)}, 
\qquad z \rightarrow 0 .
\end{equation}
Where $e_{\xi(p)(q)}$ is the polarization of   $\Phi_{\xi(p)(q)}$. 
To proceed, we shall now prove the following identity
\begin{equation}
e_{\xi(p)}=e_{(p+1)} .
\end{equation}
where, `$\xi(p)$' in the subscript means, one of the indices is $\xi$, $p$ of the indices are $+$ and the rest $s-p-1$ indices are $-$.
Let  $q$ be the number of $-$ indices we write. Now consider the `$-(p)(q)$' component 
of the first order equation (\ref{1stOrderForS}) which is given by 
\begin{equation}
\frac{g_{--}}{\sqrt{g}} (\nabla_+ \Phi _{\xi(p)(q)} - \nabla_\xi \Phi _{+(p)(q)}) = - m \Phi_{-(p)(q)} .
\end{equation}
Expanding this equation and rearranging the terms we obtain
\begin{equation}
\label{polar}
 \partial _+ \Phi_{\xi(p)(q)} - 2\sqrt{z}(1-z)\partial _z \Phi_{+(p)(q)}+
\sqrt{z} (p\Phi_{(p-1)(q+2)}-q\Phi_{(p+1)(q}) = -m\sqrt{z} \Phi_{(p)(q+1)} .
\end{equation}
In the simplifications we have also used the traceless condition (\ref{tr-s}). 
Near the horizon,  $z\rightarrow 0$, 
the first two terms of (\ref{polar}) 
go as $\sim z^{\alpha-\frac{1}{2}}$, while the other terms have the behaviour 
 $\sim z^{\alpha+\frac{1}{2}}$. 
Since the equation in (\ref{polar}) must hold   in the leading order in $z$ 
we must have the following equality
\begin{equation}
\partial _+ \Phi_{\xi(p)(q)} = 2\sqrt{z}\partial _z \Phi_{+(p)(q)} .
\end{equation}
Substituting the behaviour  near the horizon we obtain
\begin{equation}
-ik_+ e_{\xi(p)(q)} = 2\alpha e_{+(p)(q)},
\end{equation}
which implies
\begin{equation}
e_{\xi(p)(q)} = e_{+(p)(q)}. \label{coefts}
\end{equation}

Substituting
 this result in (\ref{horizon-1}), we obtain
 a recursion relation between the coefficients $e_{(p)}$ as follows 
\begin{equation}\label{recursion-1}
(s-p-1)e_{(p+2)}+(m-ik_-)e_{(p+1)}+(p+ik_+)e_{(p)}=0.
\end{equation}
This is set of $s$ equations in $s+1$ variables. Defining the `recursion matrix', $C_{jl}$ as
\begin{equation}\label{rec-11}
C^{(s)}_{jl}=(s-j-1)\delta_{j+2,l}+(m-ik_-)\delta_{j+1,l}+(j+ik_+)\delta_{j,l}.
\end{equation}
Here, $j$ runs from $s-1$ to $0$ and $l$ runs from $s$ to $0$. We can write, (\ref{recursion-1}) as
\begin{equation}\label{rec-1}
\sum_{l=s}^{0} C^{(s)}_{jl}e_{(l)} =0 \qquad \quad \text{for }j=0,1,2,\cdots,s-1.
\end{equation}
This $s\times (s+1)$ matrix $C^{(s)}$ in which the rows run from $s-1$ to $0$ and columns run from $s$ to $0$ can be written explicitly in matrix form as
\begin{equation}\label{rec-matrixform}
C^{(s)}=\begin{pmatrix}
m-ik_- &s-1+ik_+ &0              &0 &\cdots &\cdots &\cdots\\
1         &m-ik_-     &s-2+ik_+  &0 &\cdots &\cdots &\cdots \\
0			&2			&m-ik_-     &s-3+ik_+ &\cdots &\cdots &\cdots \\
\hdotsfor{7}\\
\hdotsfor{7}\\
\cdots &\cdots &\cdots &\cdots &m-ik_- &1+ik_+&0 \\
\cdots &\cdots &\cdots &\cdots &s-1 &m-ik_- &ik_+
\end{pmatrix} .
\end{equation}
We now want to write (\ref{rec-1}) in terms of the coefficients of the $\Phi_{[p]}$. Using
 (\ref{poltrans}) in (\ref{rec-1}) we get
\begin{equation}\label{rec-2}
\sum_{l=0}^{s-1} C^{(s)}_{jl} T^{(s)}_{ln}e_{[n]} =0.
\end{equation}
or, writing without the matrix labels
\begin{equation}\label{rec-3}
C^{(s)}T^{(s)}e =0,
\end{equation}
where by `$e$' we mean the column matrix containing the coefficients, $e_{[p]}$. 
To simplify the recursion relations we use the following identity which is proved
in Appendix C. 

\vspace{.5cm}\noindent
{\bf Identity 3:} 
\begin{align}\label{claim-2}
\frac{1}{2^{s-1}}\left( T^{(s-1)}C^{(s)}T^{(s)} \right) _{j,l}=\ &(2j-s+1+m+i(k_+-k_-))\delta_{j+1,l} \nn \\
&+(2j-s+1+m-i(k_+ +k_-))\delta_{j,l} .
\end{align}

\vspace{.4cm}\noindent So multiplying (\ref{rec-3}) by $T^{(s-1)}$ and using (\ref{claim-2}) we obtain\begin{align}
\left( T^{(s-1)}C^{(s)}T^{(s)} \right) _{j,l} e_{[l]}=& \ (2j-s+1+m+i(k_+-k_-))e_{[j+1]} \nn \\
&+(2j-s+1+m-i(k_+ +k_-))e_{[j]} \ = 0, \\
\text{or,} \ \  e_{[j+1]} =& - \frac{2j-s+1+m-i(k_+ +k_-)}{ 2j-s+1+m+i(k_+-k_-)} e_{[j]}.
\end{align}
Thus, all such coefficients can be written terms of $e_{[0]}$ as 
\begin{align}\label{expn-coefts}
e_{[p]}&= (-1)^p  \prod_{j=0}^{p-1} \frac{2j-s+1+m-i(k_+ +k_-)}{ 2j-s+1+m+i(k_+-k_-)}e_{[0]} .
\end{align}
Substituting this in the solution (\ref{S-soln}) we get the required solutions. 
\begin{align}\label{S-grand}
R_{[p]}  (z) &= e_{[0]} (-1)^p  \prod_{j=0}^{p-1} \frac{2j-s+1+m-i(k_+ +k_-)}{ 2j-s+1+m+i(k_+-k_-)} z^\alpha (1-z)^{\beta _{[p]} } F(a_{[p]} , b_{[p]} , c; z) .
\end{align}

\subsection{Quasinormal modes}

To obtain the quasinormal modes we need to impose the vanishing Dirichlet condition
at the boundary $z\rightarrow 1$. 
For the case $m>0$, the dominant  behaviour of the solutions near the boundary is given by
\begin{align}\label{asymp-soln}
R_{[p]}(z) &\simeq e_{[0]} (-1)^p  \prod_{j=0}^{p-1} \frac{2j-s+1+m-i(k_+ +k_-)}{ 2j-s+1+m+i(k_+-k_-)} \nn \\
& \quad \ \times z^\alpha (1-z)^{\frac{1-m-2p+s}{2}} \frac{\Gamma(c)\Gamma(a_{[p]} +b_{[p]} - c)}{\Gamma(a_{[p]})\Gamma(b_{[p]})}  .
\end{align}
To find the quasinormal  modes we need obtain the zeros  of this function. Notice that
\begin{align}
b_{[p]}&=b_{[0]}+p .
\end{align}
Therefore we have
\begin{align}
\label{expgam}
 \Gamma(b_{[p]}) 
 &=  \Gamma(b_{[0]}) \prod_{u=0}^{p-1} (b_{[0]}+u), \nn \\
 &= \frac{ \Gamma(b_{[0]})}{2^p}  \prod_{u=0}^{p-1} [2u-s+1+m-i(k_+ +k_-)].
\end{align}
Substituting  the above expression for the gamma function in (\ref{asymp-soln}), we
obtain
\begin{align}\label{final-asymp-soln}
R_{[p]}(z) &\simeq  \frac{ e_{[0]} (-2)^p}{ \prod_{j=0}^{p-1} [2j-s+1+m+i(k_+-k_-)]}  \frac{\Gamma(c)\Gamma(a_{[p]} +b_{[p]} - c)}{\Gamma(a_{[p]})\Gamma(b_{[0]})} z^\alpha (1-z)^{\frac{1-m-2p+s}{2}} .
\end{align}
It is understood here that the product in the denominator doesn't occur at all for $p=0$.

In order to impose vanishing Dirichlet conditions on the propagating components of the spin-$s$ field at infinity, we require all the above functions  which form  the components of the spin-$s$ field to vanish.  The common set of zeros for all values of $p$ is given by
\begin{equation}\label{poles-s}
a_{[s]}=-n \quad\text{and}\quad b_{[0]}=-n , \quad n =0 , 1, 2, \ldots .
\end{equation}
In terms of $k_+$ and $k_-$ these mean,
\begin{equation}\label{kk}
i(k_+ +k_-) = 2n+\hat\Delta-s, \qquad
i(k_+ -k_-) = 2n+\hat\Delta+s .
\end{equation}
Hence, the quasinormal modes of an arbitrary spin-$s$ field are
\begin{equation}
\begin{aligned}
\omega _L &=  k+2\pi T_L(k_+ +k_-) \\
&=  k-2\pi iT_L(2n + \hat \Delta -s),
\end{aligned}\qquad
\begin{aligned}
\omega _R &= -k+2\pi T_R(k_+ -k_-) \\ 
&= -k-2\pi iT_R (2n+\hat \Delta +s).
\end{aligned}
\end{equation}
These coincide precisely with the poles of the corresponding two  point function (\ref{polesl}) for the 
corresponding spin $s$ field as expected from the AdS/CFT correspondence. 
Reading out $h_L$ and $h_R$ we get $h_R -h_L=s$, the case  $h_R-h_L =-s $ will arise when 
we carry out the same analysis but with $m<0$.

\def\pol{\text{Pol($\Delta$)}}
\def\bz{\bar{z}}

\section{Higher spin 1-loop determinants}

The poles of the retarded  Green's function contain important physical information of the theory.  
As emphasized in \cite{Denef:2009kn},  they can be used to construct the one-loop determinant of the
corresponding field in the bulk. \cite{Denef:2009kn} constructed 
the one-loop determinant for scalars in asymptotically $AdS$  black holes including the BTZ black hole 
using analyticity arguments and  the information of the  quasinormal modes.
In this section we would like to use  the quasinormal modes of the spin $s$ 
fields along with the analyticity to  construct the one loop determinant of the corresponding 
spin field.  We then show that this determinant agrees with that constructed in 
\cite{David:2009xg} using group theoretic methods.

\subsection{1-loop determinant from the quasinormal spectrum}

Following \cite{Denef:2009kn} we  consider the non-rotating BTZ black hole for which the metric is given by 
\begin{equation}
\label{nonrot}
ds^2 = - (r^2 - r_+^2) dt^2 + \frac{dr^2}{(r^2 - r_+^2)} + r^2 d\phi^2 
\end{equation}
We then continue the BTZ black hole to Euclidean time together with the identification
\begin{equation}
\label{period}
 t = -i\tau,  \qquad \tau \sim \tau + \frac{1}{T}.
\end{equation}
and
\begin{equation}
T(=T_H=T_L)= \frac{r_+}{2\pi} .
\end{equation}
We are interested in evaluating the following one loop determinant of the
spin $s$ Laplacian given by
\begin{equation}
  Z_s( \hat \Delta)  = \frac{1}{2} \frac{1}{ {\rm det} ( - \nabla^2 (s) + m^2_s) } .
 \end{equation}
The basic strategy is to identify the poles of the determinant in the complex
$\hat \Delta$ space where $\hat \Delta$ is the conformal dimension of the corresponding dual operator.  This occurs whenever the wave equation of the corresponding field has a zero mode and also obeys the periodicity (\ref{period}). It is argued in \cite{Denef:2009kn}   that these zero modes are precisely  the black hole quasinormal modes. For the case of  spin $s$ field in the non-rotating  BTZ background, these modes are given by
\begin{equation} 
\omega_L  = p-2\pi i T ( 2n + \hat \Delta -s) , \qquad
 \omega_R = -p - 2\pi i T( 2n + \hat \Delta +s), \quad n =0, 1, 2,\ldots.
\end{equation}
where $p$ is the momentum along the $\theta$. It is quantized and therefore takes values in 
the set of integers
\begin{equation}
p = 0, \pm 1, \pm 2, \ldots.
\end{equation}
To proceed further, let us define the following
\begin{eqnarray}
 z_L &=&  p - 2\pi i T ( 2n + \hat\Delta -s), \qquad \; \; \  \bar z_L = p +2\pi i T( 2n + \hat \Delta -s) , \\ \nonumber
 z_R  &=& - p -2\pi i T( 2n + \hat\Delta +s), \qquad \bar z_R =  -p + 2\pi i T( 2n + \hat \Delta +s) .
\end{eqnarray}
Requiring the quasinormal modes to obey the thermal periodicity conditions given in (\ref{period}) 
results in the following equations \cite{Denef:2009kn}
\begin{eqnarray}
\label{range}
 2\pi i T( \tilde n +s)  &=&  z_L (\Delta)  ,  \qquad \tilde n \geq  0 , \\ \nonumber 
 2\pi i T( \tilde n - s)  &=& \bar z_L( \Delta ) , \qquad  \tilde n <  0  \\ \nonumber
 2\pi i T( \tilde n -s) &=& z_R (\Delta) ,  \qquad \tilde n \geq 0 ,  \\ \nonumber
 2\pi i T( \tilde n+s) &=& z_R (\Delta) , \qquad \tilde n <0   .
\end{eqnarray}
These equalities are taken to mean, that when $\hat\Delta$ is tuned to these integral  values, the 
one loop determinant exhibits poles. 
The ranges of $\tilde n$ are chosen so that the quantities
\begin{eqnarray}\nonumber
  p -2\pi i T ( 2n + \hat\Delta), \qquad \hbox{and} \quad   p +2\pi i T ( 2n + \hat\Delta),  
\end{eqnarray}
when considered together take  values $2\pi T\tilde n$ where $\tilde n$ 
assumes values in the   set of integers. 
Similarly  the range of $\tilde n$  for the case of the right-moving quasinormal modes is chosen so that the quantities,
\begin{equation} \nonumber
 -p - 2\pi i T( 2n + \hat \Delta ), \qquad \hbox{and} \quad   -p + 2\pi i T( 2n + \hat\Delta ),
\end{equation}
when considered together 
takes the  values  $2\pi i T \tilde n$ where $\tilde n$  assumes values in the set of integers.
Another requirement satisfied by the ranges in (\ref{range}) is that they reduce to that 
of  the scalar   when $s=0$. 
Though we do not have a first principle justification of the choice of these ranges we 
will show that they do indeed lead to the answer evaluated  using the group theoretic methods
given in  \cite{David:2009xg}. 
The function which is analytic in $\hat \Delta $ and has poles  at the locations (\ref{range}) is given by
\begin{align}\label{pf0}
Z_{(s)} =  \   e^{\pol} \prod _{\substack{z_L,\bz_L \\ z_R,\bz_R}} \Bigg{[} &  
\left( s+\frac{iz_L}{2\pi T} \right)^{-1} 
\left( -s+\frac{iz_R}{2\pi T} \right)^{-1}
 \nonumber \\
\times   &\; \prod _{n\ge s} \left( n+\frac{iz_L}{2\pi T} \right)^{-1}  \left( n-\frac{i\bz_L}{2\pi T} \right)^{-1} \nn \\
&  \prod _{n\ge -s} \left( n+\frac{iz_R}{2\pi T} \right)^{-1}  \left( n-\frac{i\bz_R}{2\pi T} \right)^{-1}  \Bigg{]},
\end{align}
where $\pol$ is a non-singular holomorphic function of $\hat \Delta$ and can be determined by examining the $\hat \Delta \rightarrow \infty$ behaviour.  The product over $z_L, \bar z_L, z_R, \bar z_R$
mean the product over $p, n, \tilde n$  occurring in the definition of these variables. 
Using the following relations
\begin{equation}
 i( z_L + \bar z_R) =  -2\pi T( 2s) , \qquad i ( \bar z_L  +  z_R) = 2\pi T (2s)  .
\end{equation}
we can write the one loop determinant  as
\begin{align}\label{pf}
Z_{(s)} =  \   e^{\pol} \prod _{\substack{z_L,\bz_L \\ z_R,\bz_R}} \Bigg{[} &  \left( s+\frac{iz_L}{2\pi T} \right)^{1/2} 
\left( s-\frac{i\bz_L}{2\pi T} \right)^{1/2} \left( -s+\frac{iz_R}{2\pi T} \right)^{1/2} \left( -s-\frac{i\bz_R}{2\pi T} \right)^{1/2} \nonumber \\
\times   &\; \prod _{n\geq s} \left( n+\frac{iz_L}{2\pi T} \right)^{-1}  \left( n-\frac{i\bz_L}{2\pi T} \right)^{-1} \nn \\
\times &  \prod _{n\geq -s} \left( n+\frac{iz_R}{2\pi T} \right)^{-1}  \left( n-\frac{i\bz_R}{2\pi T} \right)^{-1}  \Bigg{]}.
\end{align}
Note that according to the analysis of \cite{Denef:2009kn} the expression in (\ref{pf}) is the partition 
function for a complex field and therefore one needs to take a square root for a 
real field. However for our higher spin $s>0$, we have two modes, corresponding to 
$h_R-h_L=\pm s$. Taking into account both these modes results in the expression given in (\ref{pf}). 
We now substitute in the values of the left and right quasinormal modes and perform the
same manipulations as in \cite{Denef:2009kn} to obtain
\begin{eqnarray}
Z_{(s)} &=&  \   e^{\pol}  \prod _{N\geq 0, p} \Bigg{[}  \left(( 2N+\hat \Delta)^2 +
\frac{p^2}{(2\pi T)^2} \right)   \nn \\
&& \qquad \qquad \qquad \  \times \prod _{n\geq 0} \left((n+ 2N+\hat \Delta)^2 +\frac{p^2}{(2\pi T)^2} \right)^{-2} \Bigg{]}   .
\end{eqnarray}
Taking logarithms on both sides of the equation lead to
\begin{align}
 - \log Z_{(s)}  &=  - \pol + 2 \sum_{n\geq 0, N\geq 0, p} \log \left((n+ 2N+\hat \Delta)^2 +\frac{p^2}{(2\pi T)^2} \right) \nn \\
&\qquad\qquad\quad -  \sum _{N\geq 0,p} \log 
\left(( 2N+\hat\Delta)^2 +\frac{p^2}{(2\pi T)^2} \right)  \notag \\ 
	 &=  - \pol + 2 \sum_{n> 0, N\geq 0, p} \log \left((n+ 2N+\hat\Delta)^2 +\frac{p^2}{(2\pi T)^2} \right)  \nn \\
&\qquad\qquad\quad +  \sum _{N\geq 0,p} \log \left(( 2N+\hat \Delta)^2 +\frac{p^2}{(2\pi T)^2} \right)  \notag \\
     &=  - \pol +  \sum _{\kappa \geq 0, p} (\kappa + 1) \log \left((\kappa +\hat\Delta)^2 +\frac{p^2}{(2\pi T)^2} \right)  .
\end{align}
Here  the sums over $n$ and $N$ were combined and written as a sum 
over $\kappa  =n+2N =0, 1, \ldots. $ The factor $\kappa +1$ accounts for the multiplicity. 
 We have also made use of $\log (a+ib) + \log (a-ib) = \log (a^2+b^2)$.
We then extract out the divergent terms which are absorbed in $\pol$ and 
 make use of the identity 
\(  \sum _{p\geq1} \log \left( 1+ \frac{x^2}{p^2}\right) =
 \log \frac{\sinh  \pi x}{\pi x} = \pi x -\log (\pi x) + \log (1- e^{-2\pi x}) \) to obtain
\begin{equation}\label{pf3}
\log Z_{(s)} = \pol +2 \log \prod _{\kappa \geq 0} (1- q^{-\kappa +\hat \Delta})^{-(\kappa + 1 )}.
\end{equation}
Here
\begin{equation}
q=e^{2\pi i \tilde{\tau}} \ , \quad \quad \tilde{\tau}=2\pi i T  .
\end{equation}
Now we can use the same  argument as in \cite{Denef:2009kn} to determine $\pol$. 
Taking $\hat \Delta\rightarrow \infty$, the partition function should reduce to that of the BTZ which 
is locally identical to that of $AdS_3$. This determines $\pol$ to be a function proportional to the 
volume of the Euclidean BTZ black hole. We will not write this explicitly since 
we do not  require it in the subsequent discussion. 

\subsection{1-loop determinant  from the heat kernel}

We will now show that the term in the one loop partition function (\ref{pf3}) 
 determined form the quasinormal modes agrees with that constructed from the 
 heat kernel of the spin $s$ field.  The trace of the heat kernel for the spin $s$  Laplacian on thermal $AdS_3$  is given by
\cite{David:2009xg},
\begin{equation}
{\rm Tr} ( e^{-t \nabla_{(s)}}) = 
K^{(s)}(\tau,\bar{\tau};t) = \sum ^{\infty} _{n=1} \frac{\tau _2}{\sqrt{4\pi t}|\sin \frac{n\tau}{2}|^2}\cos (sn\tau_1) e^{-\frac{n^2 \tau _2^2}{4t}} e^{-(s+1)t}.
\end{equation}
Note that in the above expression we have suppressed the term which is proportional to the 
volume of  the $AdS_3$.  Here $\tau$ is related to the 
temperature of the Euclidean  non-rotating BTZ by,
\begin{equation}
\tau = \frac{i}{2\pi T}.
\end{equation}
We will substitute this value of $\tau$ towards the end 
of our analysis. 
The 1-loop  determinant  is then  given by
\begin{equation}
-\frac{1}{2}\log (\det (-\nabla ^2 + m_s^2 ) )=
\frac{1}{2}\int _0^{\infty} \frac{dt}{t} e^{-m^2_s t} K^{(s)}(\tau,\bar{\tau};t) .
\end{equation}
where $m_s^2$ is defined in (\ref{defms}).  
Substituting the values of $m_s$ and using the definition of $\hat\Delta$  given 
in (\ref{conf-dim}) we obtain
\begin{equation}
 e^{-m^2_s t} K^{(s)} (\tau,\bar{\tau};t)=\frac{1}{2} \sum _{n=1} ^\infty \frac{\tau _2}{\sqrt{4\pi t}|\sin \frac{n\tau}{2}|^2}\cos (sn\tau_1) e^{-\frac{n^2\tau_2^2}{4t}}e^{-(\hat\Delta -1)^2t}.
\end{equation}    
The integration over $t$ can be easily performed using
\begin{equation}
\frac{1}{\sqrt{4\pi}} \int_0^\infty \frac{dt}{t^{3/2} }e^{-\frac{\alpha^2}{4t} - \beta^2 t} = \frac{1}{\alpha} e^{-\alpha\beta}.
\end{equation}
we then obtain
\begin{equation}
\begin{aligned}\label{q3m}
-\frac{1}{2} \log (\det (-\nabla^2+m_s^2)) &=
\frac{1}{2} \sum _{n=1} ^\infty \frac{\cos (sn\tau_1)}{n|\sin \frac{n\tau}{2}|^2}  e^{-n\tau_2(\Delta -1)} \\
&= \sum _{n=1} ^\infty \frac{1}{n} \frac{(q^{sn}+\bar q^{sn})}{|1-q^n|^2}q^{(\Delta-s)n} \\
&= \sum _{n=1} ^\infty \frac{2}{n} \frac{q^{\Delta n}}{(1-q^n)^2} \\
&= -2 \log \prod_{m=0} ^\infty (1-q^{m+\Delta} )^{m+1} .
\end{aligned} 
\end{equation}
where
\begin{equation}
q = e^{2\pi i \tau}  .
\end{equation}
In the third step we have used the fact that $\tau$ is purely imaginary for the case of 
the non-rotating BTZ black hole which results in $q= \bar q$. 
Therefore we obtain
\begin{equation}\label{qf1/2}
\log Z_s = -\frac{1}{2} \log (\det (-\nabla^2+m_s^2)) = -2 \log \prod_{m=0} ^\infty (1-q^{m+\Delta} )^{m+1} .
\end{equation}
Comparing the (\ref{pf3}) and (\ref{qf1/2}) we see that the two expressions
indeed agree on the performing the modular transformation,
\begin{equation}
\tilde \tau = - \frac{1}{\tilde \tau}.
\end{equation}
which is the expected relation between the one-loop determinants on 
Euclidean BTZ and thermal $AdS_3$.

\section{Conclusions}

We have solved the wave equations for massive higher integer spin fields in the BTZ background. 
To obtain the quasinormal modes we have focused only on the ingoing modes at the horizon, but the 
analysis can easily be carried out for the outgoing modes. 
This will lead to the complete set of modes for the  massive higher spin field which is the 
starting point to quantize the field in this background. 
It will be useful to carry out this analysis in detail to study quantum properties of the black hole 
like Hawking radiation. 
From our discussion of the wave equations, it seems that other quantities like the bulk to boundary 
propagator for massive higher spin fields in $AdS_3$ can also be solved. 
These are important tools to study the $AdS_3/CFT_2$ correspondence for massive  higher spin fields 
and useful  to obtain them. Another obvious direction to extend this work is to consider the case 
of the fermionic higher spin fields. 

At present there are only a few examples of black hole backgrounds  known
in which the string  world sheet theory can be  quantized.  
The 2d black hole \cite{Mandal:1991tz, Witten:1991yr}
and the BTZ background with NS flux \cite{Maldacena:2000kv} are the well known cases.
The analysis of this paper and the fact the string motion in the BTZ black hole is integrable 
\cite{David:2011iy} suggests that it might be possible to quantize the string in this background
which is clearly an interesting direction to pursue.

\acknowledgments

We wish to thank  Rajesh Gopakumar,  Gautam Mandal, 
Shiraz Minwalla, Spenta Wadia and for useful discussions regarding the 
general properties of  quasinormal modes in higher spin theories. 
We also thank B.  Ananthanarayan,  Kavita Jain and Sujit Nath for helpful discussions 
on some technical aspects of this problem. 
 We thank the
 the International Centre of Theoretical Sciences of the TIFR for organizing a stimulating
meeting at IISc, Bangalore for facilitating this discussion. The work of 
J.R.D is partially supported by the Ramanujan fellowship DST-SR/S2/RJN-59/2009.

\appendix

\section{The vector case}
The spin-1 case is a special case which does not fall in our 
general description of higher spin fields. This is because we can impose the tracelessness condition for fields with $s\geq2$. Nevertheless, in this section we shall use the similar methods developed for the graviton and 
for higher spin cases  to calculate the quasinormal modes of the $s=1$ field. 
We use the Chern-Simons equations of motion for a massive gauge field
\begin{equation}
\label{s1first}
 \epsilon_{\mu}^{\ \alpha\beta} \partial_\alpha A_\beta = -m A_\mu .
\end{equation}
Using this we obtain the following equations for the components
\begin{eqnarray}
\label{compv1} \frac{g_{--}}{\sqrt{-g}} (\partial_+A_z - \partial_z A_+) =  -m  A_- ,\\
\label{compv2}  \frac{g_{++}}{\sqrt{-g}} (\partial_z A_- - \partial_-A_z ) =  -m A_+ .
\end{eqnarray}
It can be shown that the above equation is equivalent to 
second order massive spin equations for the vector field, with \(m=M\). 

\subsubsection*{Reduction of the vector Laplacian to the scalar Laplacian}
We will now show that the spin-1 Laplacian acting on components $A_+$ and $A_-$ 
can be reduced to the scalar Laplacian
\begin{align}
 \nabla^2 A_+ &= \frac{1}{\sqrt{g}} \partial_\mu( \sqrt{g} g^{\mu\nu} \partial_\nu A_+) 
- \frac{1}{\sqrt{g}} \partial_\mu ( \sqrt{g} g^{\mu\nu} \Gamma_{\nu+}^\sigma ) A_\sigma
- 2 \Gamma^\sigma_{\mu+}g^{\mu\nu} \partial_\nu A_\sigma
+\Gamma^\sigma_{\mu+} g^{\mu\nu} \Gamma_{\nu\sigma}^\alpha A_\alpha \nonumber \\
&= \frac{1}{\sqrt{g}} \partial_\mu( \sqrt{g} g^{\mu\nu} \partial_\nu A_+) 
- \frac{1}{\sqrt{g}} \partial_\xi ( \sqrt{g} g^{\xi\xi} \Gamma_{\xi +}^+) A_+
- 2 \Gamma^\xi_{++}g^{++} \partial_+ A_\xi - 2 \Gamma^+_{\xi +}g^{++} \partial_\xi A_+ \nonumber \\
&\qquad \qquad +\Gamma^\xi _{++} g^{++} \Gamma_{+z}^+ A_+ +\Gamma^\xi_{\xi +} g^{\xi\xi} \Gamma_{+z}^+ A_+ \nonumber \\
&= \Delta A_- -2A_+ +2\coth \mu (\pd_+ A_\xi -\pd_\xi A_+ )+(\Gamma^\xi_{++} g^{++} +\Gamma^\xi_{\xi +} )\Gamma_{+\xi}^+ A_+  \nonumber \\
&= \Delta A_- -2A_+ -2mA_- .
\end{align}
Here we  have used (\ref{compv2}) and (\ref{c1}) to obtain  the last line in the above equation.  
Now let's consider the laplacian acting on $A_-$. We get
\begin{align}
 \nabla^2 A_- &= \frac{1}{\sqrt{g}} \partial_\mu( \sqrt{g} g^{\mu\nu} \partial_\nu A_-) 
- \frac{1}{\sqrt{g}} \partial_\mu ( \sqrt{g} g^{\mu\nu} \Gamma_{\nu-}^\sigma ) A_\sigma
- 2 \Gamma^\sigma_{\mu-}g^{\mu\nu} \partial_\nu A_\sigma
+\Gamma^\sigma_{\mu-} g^{\mu\nu} \Gamma_{\nu\sigma}^\alpha A_\alpha \nonumber \\
&= \frac{1}{\sqrt{g}} \partial_\mu( \sqrt{g} g^{\mu\nu} \partial_\nu A_-) 
- \frac{1}{\sqrt{g}} \partial_\xi ( \sqrt{g} g^{\xi\xi} \Gamma_{\xi-}^-) A_-
- 2 \Gamma^\xi_{--}g^{--} \partial_- A_\xi- 2 \Gamma^-_{\xi-}g^{--} \partial_\xi A_- \nonumber \\
&\qquad \qquad +\Gamma^\xi_{--} g^{--} \Gamma_{-\xi}^- A_- +\Gamma^\xi_{\xi-} g^{\xi\xi} \Gamma_{-\xi}^- A_- \nonumber \\
&= \Delta A_- -2A_- +2\tanh \mu (\pd_- A_\xi -\pd_\xi A_- )+(\Gamma^\xi_{--} g^{--} +\Gamma^\xi_{\xi-}  )\Gamma_{-\xi}^- A_-  \nonumber \\
&= \Delta A_- -2A_- -2mA_+ .
\end{align}
where we have used 
(\ref{compv1}) and (\ref{c1}) to obtain the  last equality.  
To summarize for the spin-1 case, the spin one Laplacian reduces to the scalar Laplacian as follows
\begin{align}
\label{spin1s}
\nabla ^2 A_+ &= \Delta A_+ -2mA_- - 2A_+ ,\\ \nonumber
\nabla ^2 A_- &= \Delta A_- -2A_- -2mA_+ .
\end{align} 
with
\begin{equation}
\Delta A_{\alpha} =  \frac{1}{\sqrt{g}} \partial_\mu( \sqrt{g} g^{\mu\nu} \partial_\nu A_\alpha) .
\end{equation}

\subsection*{Solutions of the spin-1 components}

The equations in (\ref{spin1s})
can be decoupled  by considering the combinations $A_{1,2}=A_+ \pm A_-$. 
We then have
\begin{align} \label{vec-dec-1}
\nabla^2 A_1 &= (\Delta -2m -2) A_1 , \\  \label{vec-dec-2}
\nabla^2 A_2 &= (\Delta +2m -2) A_2 .
\end{align}
The second order equation of motion satisfied by the spin one field is given by
\begin{equation}
(\nabla ^2 -m^2+2)A_\mu =0 .
\end{equation}
Using (\ref{vec-dec-1}) and (\ref{vec-dec-2}) we obtain the following equations of 
motion for the components $A_1, A_2$. 
\begin{align}
\label{dcoups1}
(\Delta - (m+1)^2 +1)A_1 &=0, \\ \nonumber
(\Delta - (m-1)^2 +1)A_2 &=0 .
\end{align}
We now substitute the ansatz
\begin{align}
A_1 = e_1 e^{-i(k_+x^+ + k_-x^-)} R_1 (\xi), \\ \nonumber
A_2 = e_2 e^{-i(k_+x^+ + k_-x^-)} R_2 (\xi) .
\end{align}
into the equations (\ref{dcoups1}).
Expanding out the scalar Laplacian result in  the following equations
\begin{align}
z(1-z)\frac{d^2 R_{1} }{dz^2} + (1-z)\frac{dR_{1} }{dz} + \Big{[} \frac{k_+^2}{4z} -  \frac{k_-^2}{4} - \frac{(m+1)^2 -1 }{4(1-z)}\Big{]}  R_{1} &= 0  , \\ \nonumber
z(1-z)\frac{d^2 R_{2} }{dz^2} + (1-z)\frac{dR_{2} }{dz} + \Big{[} \frac{k_+^2}{4z} -  \frac{k_-^2}{4} - \frac{(m-1)^2 -1 }{4(1-z)}\Big{]}  R_{2} &= 0 .
\end{align}
where $z=\tanh^2 \xi$. The solution that obeys ingoing boundary conditions at the horizon is given by
\begin{align}\label{hypgeo-vector}
R_{1}  (z) &=e_{1}  z^\alpha (1-z)^{\beta _{1} } F(a_{1} , b_{1} , c; z), \\ \nonumber
R_{2}  (z) &=e_{2}  z^\alpha (1-z)^{\beta _{2} } F(a_{2} , b_{2} , c; z) .
\end{align}
where 
\begin{eqnarray}
\alpha &=& \frac{-ik_+}{2}, \qquad c = 1+ 2\alpha, \\ \nonumber
\beta _{1}  &=& \frac{m}{2} +1, \qquad
\beta _{2}  = \frac{m}{2},  \\ \nonumber
a_{1,2}  &=& \frac{k_+ - k_-}{2i} + \beta _{1,2},  \qquad
b_{1,2}  = \frac{k_+ + k_-}{2i} + \beta _{1,2}  .
\end{eqnarray}

The polarization constants $e_1, e_2$ are not independent by are determined by the 
first order equation (\ref{s1first}).  Since they are constants to determine it is sufficient to 
examine the behaviour of the solutions near the horizon $z\rightarrow0$. This is given by
\begin{equation}
A_{1,2} \rightarrow  e_{1,2} z^\alpha e^{-i(k_+x^+ + k_-x^-)}, \qquad z\rightarrow 0.
\end{equation}
Therefore we also have the relation,
\begin{equation}
 A_\pm  \rightarrow (e_1 \pm e_2) z^\alpha e^{-i(k_+x^+ + k_-x^-)}, \qquad z\rightarrow 0 .
\end{equation}
The $A_\xi$ component can be obtained from (\ref{s1first}), which is given by
\begin{eqnarray}
A_{\xi} &=& \frac{1}{m \cosh \xi \sinh \xi} \partial_{[+}A_{-]} \\ \nonumber
			&=& \frac{i(1-z)}{2m\sqrt{z}} (k_-A_+ - k_+A_-). 
\end{eqnarray}
The above equation leads to the following  near horizon behaviour of $A_\xi$ 
\begin{eqnarray}\nonumber
A_{\xi} &\rightarrow& 
 \frac{i}{2m} ((k_- -k_+)e_1 + (k_- +k_+)e_2 )z^{\alpha -\frac{1}{2}} e^{-i(k_+x^+ + k_-x^-)}
, \quad z\rightarrow 0, \\ \label{a-xi}
&=& e_\xi z^{\alpha -\frac{1}{2}} e^{-i(k_+x^+ + k_-x^-)} .
\end{eqnarray}
Let's now consider the $(-)$ component of the first order equation 
\begin{eqnarray}
-mA_- &=& \epsilon_{-}^{\ \alpha\beta} \partial _\alpha A_\beta \\ \nonumber
           &=& -2(1-z)\partial _z A_+ + \frac{1}{\sqrt{z}} \partial _+ A_\xi .
\end{eqnarray}
Examining the above equation near the 
 horizon $z\rightarrow0$, the terms on the right tend to   $\sim z^{\alpha-1}$. Thus they are dominant 
compared to that on the left  ($\sim z^{\alpha}$).   Substituting the behaviour of $A_+$ and $A_\xi$ near 
the horizion we obtain the relation  
\begin{equation}
 -ik_+ e_{\xi} = \alpha (e_1 +e_2) .
\end{equation}
which results in 
\begin{equation}
 e_{\xi} =\frac{e_1 +e_2}{2} .
\end{equation}
Using the above equation in  (\ref{a-xi}) we obtain
\begin{align}
\frac{e_1}{e_2} = \frac{i(k_+ +k_-)-m}{i(k_+ -k_-)+m}.
\end{align}
Finally substituting this in the solution (\ref{hypgeo-vector}) we obtain the following solutions
\begin{align}
\label{solns1}
R_{1}  (z) &=e_{2}  \frac{i(k_+ +k_-)-m}{i(k_+ -k_-)+m}  z^\alpha (1-z)^{\beta _{1} } F(a_{1} , b_{1} , c; z), \\
R_{2}  (z) &=e_{2}  z^\alpha (1-z)^{\beta _{2} } F(a_{2} , b_{2} , c; z) .
\end{align}

\subsection*{Quasinormal modes}

Quasinormal modes are obtained by imposing vanishing Dirichlet boundary conditions 
at the boundary $z\rightarrow 1$.  The leading behaviour of (\ref{solns1}) at the boundary for $m>0$ is given by 
\begin{align}
R_{1}  (z) &\simeq e_{2}  \frac{i(k_+ +k_-)-m}{i(k_+ -k_-)+m}   \frac{\Gamma(c)\Gamma(a_1 + b_1 -c)}{\Gamma(a_1)\Gamma(b_1)}  (1-z)^{-\frac{m}{2}}  \nn \\
&=  -\frac{ e_{2}}{i(k_+ -k_-)+m}   \frac{\Gamma(c)\Gamma(a_1 + b_1 -c)}{\Gamma(a_1)\Gamma(b_1 -1 )}  (1-z)^{-\frac{m}{2}} , \\
R_{2}  (z) &\simeq e_{2}  (1-z)^{1-\frac{m}{2}} \frac{\Gamma(c)\Gamma(a_2 + b_2 -c)}{\Gamma(a_2)\Gamma(b_2)}  (1-z)^{1-\frac{m}{2}} .
\end{align}
The common set of zeros of the above functions are given by
\begin{equation}
b_1 -1 = b_2 = -n , \qquad a_1=-n., \qquad n = 0, 1, ,2, \cdots .
\end{equation}
In terms of $k_+$ and $k_-$ these are
\begin{equation}
i(k_+ + k_-) = 2n+\hat \Delta -2 \ , \qquad i(k_+ - k_-) = 2n+\hat \Delta +2 .
\end{equation}
where $\hat \Delta = 1 +m$.  Expressing $k_+$ and $k_-$ in terms of frequency and momenta 
we see that the quasinormal modes for the gauge field are given by
\begin{eqnarray}
\omega _L &=& k+2\pi T_L(k_+ +k_-), \qquad  \omega _R = -k+2\pi T_R(k_+ -k_-) \\ \nonumber
				&=&  k-2\pi iT_L(2n + \hat\Delta -1), \qquad 
				= -k-2\pi iT_R (2n+ \hat\Delta +1).
\end{eqnarray}
These modes coincide with the poles of the two point function (\ref{polesl}) for the $s=1$ case. 
Note that we have obtained the situation with $h_R-h_L =1$. If we perform the same analysis with 
$m<0$ we will obtain $h_R-h_L =-1$.

\section{The spin 3 case}
In this section we shall find the quasinormal modes of the spin-3 field.
This is to demonstrate  the methods  developed for the arbitrary-$s$ calculations explicitly.
The first order equation for this case is
\begin{equation}
\epsilon_\mu ^{\ \alpha \beta} \nabla _\alpha \phi_{\beta \nu \eta} = -m \phi_{\mu \nu \eta}.
\end{equation}

\subsection*{Reduction of the spin-3 Laplacian to the scalar Laplacian}
From (\ref{laplacian-on-s}) we get the
following  expressions  for the Laplacian acting on various components of 
$\phi_{\mu\nu\eta}$ in terms of the scalar Laplacian. These are
\begin{align}
\nabla ^2 \p _{+++} &=  \Delta \p _{+++} -6\p_{+++}  - 6m \p_{-++} - 6\p_{--+} , \\
\nabla ^2 \p _{---} &=  \Delta \p _{---} -6\p_{---}  - 6m \p_{+--} - 6\p_{++-} ,\\
\nabla ^2 \p _{++-} &=  \Delta \p _{++-} -6\p_{++-}  - 4m \p_{-+-} -2m \p_{+++} -2 \p_{---} - 4\p_{++-} ,\\
\nabla ^2 \p _{--+} &=  \Delta \p _{--+} -6\p_{--+}  - 4m \p_{+-+} -2m \p_{---} -2 \p_{+++} - 4\p_{--+} .
\end{align}
Rewriting  the above equation  in the form of a  matrix as in (\ref{lap-scalar&mass}) results in
\begin{align}
\label{m3mat}
\nabla^2 \phi_{(p)} &= \Delta \phi_{(p)}  + M^{(3)}_{pq}\phi_{(q)}  \\
\nabla ^2 \begin{pmatrix}
\p_{+++} \\
\p_{++-} \\
\p_{+--} \\
\p_{---} 
\end{pmatrix} &= \Delta  \begin{pmatrix}
\p_{+++} \\
\p_{++-} \\
\p_{+--} \\
\p_{---} 
\end{pmatrix} + \begin{pmatrix}
-6 &-6m &-6 &0 \\
-2m &-10 &-4m &-2 \\
-2 &-4m &-10 &-2m \\
0 &-6 &-6m &-6
\end{pmatrix}\begin{pmatrix}
\p_{+++} \\
\p_{++-} \\
\p_{+--} \\
\p_{---} 
\end{pmatrix} .
\end{align}

\subsection*{Diagonalization of the mass matrix}

We now require to diagonalize the mass matrix
The transformation matrix that diagonalizes the 
$M^{(3)}$ in (\ref{m3mat}) can be obtained from (\ref{T-expression}). This is given by,
\begin{equation}
T^{(3)}= \begin{pmatrix}
1 &3 &3 &1 \\
1 &1 &-1 &-1 \\
1 &-1 &-1 &1 \\
1 &-3 &3 &-1
\end{pmatrix}.
\end{equation}
Then  the linear combination for the components of the field for 
which the equations decoupled is
\begin{equation}
\begin{pmatrix}
\phi_{111} \\
\phi_{112} \\
\phi_{122} \\
\phi_{222} 
\end{pmatrix} =  \begin{pmatrix}
1 &3 &3 &1 \\
1 &1 &-1 &-1 \\
1 &-1 &-1 &1 \\
1 &-3 &3 &-1
\end{pmatrix}\begin{pmatrix}
\p_{+++} \\
\p_{++-} \\
\p_{+--} \\
\p_{---} 
\end{pmatrix} .
\end{equation}
It can then be seen by explicit matrix multiplication that
\begin{equation}
T^{(3)}M^{(3)}(T^{(3)})^{-1}=\begin{pmatrix}
-12-6m &0 &0 &0 \\
0 &-4-2m &0&0 \\
0 &0 &-4+2m &0 \\
0 &0 &0 &-12+6m
\end{pmatrix} .
\end{equation}
This verifies the identity (\ref{claim-1}) for $s=3$.
Therefore  we obtain
\begin{equation}
\nabla^2 
\begin{pmatrix}
\phi_{111} \\
\phi_{112} \\
\phi_{122} \\
\phi_{222} 
\end{pmatrix} = \begin{pmatrix}
\Delta-12-6m &0 &0 &0 \\
0 &\Delta-4-2m &0&0 \\
0 &0 &\Delta-4+2m &0 \\
0 &0 &0 &\Delta-12+6m
\end{pmatrix}\begin{pmatrix}
\phi_{111} \\
\phi_{112} \\
\phi_{122} \\
\phi_{222} 
\end{pmatrix}.
\end{equation}
The second order equation satisfied by the the spin 3 field can be read out from 
 (using (\ref{2ndOrder})). This is given by
\begin{equation}
(\nabla^2 - m^2 + 4) \phi _{\mu\nu\eta} = 0 .
\end{equation}
Thus the component  $\phi_{111}, \phi_{112}, \phi_{122}$ and $\phi_{222}$ satisfy the following equations
\begin{align}
(\Delta -(m+3)^2 +1)\phi_{111} &= 0 ,\\ \nonumber
(\Delta -(m+1)^2 +1)\phi_{112} &= 0 ,\\ \nonumber
(\Delta -(m-1)^2 +1)\phi_{122} &= 0 ,\\ \nonumber
(\Delta -(m-3)^2 +1)\phi_{111} &= 0 .
\end{align}
Written out explicitly in terms of the coordinates with $z=\tanh^2 \xi$  and using the ansatz $\phi_{\mu\nu\eta}=e^{-i(k_+ x^+ + k_-x^-)}R_{\mu\nu\eta}$, the above equations  reduce to the following
\begin{equation}\label{hyp-3}
\begin{aligned}
z(1-z)\frac{d^2 R_{111}}{dz^2} + (1-z)\frac{dR_{111}}{dz} + \Big{[} \frac{k_+^2}{4z} -  \frac{k_-^2}{4} - \frac{(m+3)^2 -1 }{4(1-z)}\Big{]}  R_{111} &= 0, \\ 
z(1-z)\frac{d^2 R_{112}}{dz^2} + (1-z)\frac{dR_{112}}{dz} + \Big{[} \frac{k_+^2}{4z} -  \frac{k_-^2}{4} - \frac{(m+1)^2 -1 }{4(1-z)}\Big{]}  R_{112} &= 0, \\ 
z(1-z)\frac{d^2 R_{122}}{dz^2} + (1-z)\frac{dR_{122}}{dz} + \Big{[} \frac{k_+^2}{4z} -  \frac{k_-^2}{4} - \frac{(m-1)^2 -1 }{4(1-z)}\Big{]}  R_{122} &= 0, \\
z(1-z)\frac{d^2 R_{222}}{dz^2} + (1-z)\frac{dR_{222}}{dz} + \Big{[} \frac{k_+^2}{4z} -  \frac{k_-^2}{4} - \frac{(m-3)^2 -1 }{4(1-z)}\Big{]}  R_{222} &= 0 . 
\end{aligned}
\end{equation}

\subsection*{Solutions of the spin-3 components}
The solutions of (\ref{hyp-3}) that obeys ingoing boundary conditions at the horizon are
\begin{align}
R_{111} (z) &=e_{111} z^\alpha (1-z)^{\beta _{111}} F(a_{111}, b_{111}, c; z), \\
R_{112} (z) &=e_{112} z^\alpha (1-z)^{\beta _{112}} F(a_{112}, b_{112}, c; z) ,\\
R_{122} (z) &=e_{122} z^\alpha (1-z)^{\beta _{122}} F(a_{122}, b_{122}, c; z) ,\\
R_{222} (z) &=e_{222} z^\alpha (1-z)^{\beta _{222}} F(a_{222}, b_{222}, c; z) .
\end{align}
where,
\begin{eqnarray}
\alpha &=& \frac{-ik_+}{2} , \qquad c = 1+ 2\alpha, \\ \nonumber
\beta _{111} &=& \frac{1}{2} (m+4), \quad 
\beta _{112} = \frac{1}{2} (m+2), \qquad 
\beta _{122} = \frac{1}{2} m , \quad 
\beta _{222} = \frac{1}{2} (m-2) \\ \nonumber
a_{ijl}  &=& \frac{k_+ - k_-}{2i} + \beta _{ijl}, \qquad
b_{ijl} = \frac{k_+ + k_-}{2i} + \beta _{ijl}  .
\end{eqnarray}

\subsection*{Fixing the polarization constants}

After investigating the behaviour near the horizon we obtain  
from (\ref{rec-1})  a set of recursion relations for the polarization constants. This is of the form $C^{(3)}\cdot e=0$. Writing them out explicitly we get
\begin{equation}\label{3-rec}
\begin{pmatrix}
m-ik_- &2+ik_- &0 &0 \\
1 &m-ik_- &1+ik_- &0 \\
0 &2 &m-ik_- &ik_+
\end{pmatrix}\begin{pmatrix}
e_{+++} \\
e_{++-} \\
e_{+--} \\
e_{---} 
\end{pmatrix} =0 .
\end{equation}
It can then be verified that
\begin{align}
\frac{1}{2^{s-1}}&T^{(2)}C^{(3)}T^{(3)} \nn \\
&={\footnotesize \begin{pmatrix}
2+m+i(k_+ - k_-) &2+m-i(k_+ + k_-) &0 &0 \\
0 &m+i(k_+ - k_-) &m-i(k_+ + k_-) &0 \\
0 &0 &-2+m+i(k_+ - k_-) &-2+m-i(k_+ + k_-)
\end{pmatrix}}.
\end{align}
This verifies (\ref{claim-2}) for $s=3$. 
The above relation then gives a simple
 recursion relation between two $e_{[p]}$ from which we can obtain all coefficients in terms of $e_{[0]}$. 
Using (\ref{expn-coefts}) we have
\begin{align}
e_{111} &= - \prod_{j=0}^{2} \frac{2j-2+m-i(k_+ + k_-)}{2j-2+m+i(k_+ - k_-)} e_{222} \\
e_{112} &=  \prod_{j=0}^{1} \frac{2j-2+m-i(k_+ + k_-)}{2j-2+m+i(k_+ - k_-)} e_{222} \\
e_{122} &= -  \frac{-2+m-i(k_+ + k_-)}{-2+m+i(k_+ - k_-)} e_{222}
\end{align}
Thus, the final form of the solutions are
\begin{align}
R_{111} (z) &=-e_{222}  \prod_{j=0}^{2} \frac{2j-2+m-i(k_+ + k_-)}{2j-2+m+i(k_+ - k_-)} z^\alpha (1-z)^{\beta _{11}} F(a_{111}, b_{111}, c; z) \\
R_{112} (z) &=e_{222}  \prod_{j=0}^{1} \frac{2j-2+m-i(k_+ + k_-)}{2j-2+m+i(k_+ - k_-)} z^\alpha (1-z)^{\beta _{112}} F(a_{112}, b_{112}, c; z) \\
R_{122} (z) &=-e_{222}  \frac{-2+m-i(k_+ + k_-)}{-2+m+i(k_+ - k_-)}  z^\alpha (1-z)^{\beta _{122}} F(a_{122}, b_{122}, c; z) \\
R_{222} (z) &=e_{222} z^\alpha (1-z)^{\beta _{222}} F(a_{222}, b_{222}, c; z) 
\end{align}

\subsection*{Quasinormal modes}
We can now obtain the quasinormal modes by imposing vanishing Dirichlet conditions
at the boundary.  For $m>0$ the 
leading  behaviour of these solutions near the boundary $z\rightarrow 1$ is given by
\begin{align}
R_{111} (z) &\simeq -e_{222}  \prod_{j=0}^{2} \frac{2j-2+m-i(k_+ + k_-)}{2j-2+m+i(k_+ - k_-)} z^\alpha (1-z)^{-\frac{m}{2}-1} \frac{\Gamma(c)\Gamma(a_{111}+b_{111}-c)}{\Gamma(a_{111})\Gamma(b_{111})} \nn \\
&=  -  \frac{8e_{222} }{\prod_{j=0}^{2} [2j-2+m+i(k_+ - k_-)]} z^\alpha (1-z)^{-\frac{m}{2}-1} \frac{\Gamma(c)\Gamma(a_{111}+b_{111}-c)}{\Gamma(a_{111})\Gamma(b_{111}-3)} ,\\
R_{112} (z) &\simeq e_{222}  \prod_{j=0}^{1} \frac{2j-2+m-i(k_+ + k_-)}{2j-2+m+i(k_+ - k_-)} z^\alpha (1-z)^{-\frac{m}{2}} \frac{\Gamma(c)\Gamma(a_{112}+b_{112}-c)}{\Gamma(a_{112})\Gamma(b_{112})} \nn \\
&=  \frac{4e_{222} }{ \prod_{j=0}^{1} [ 2j-2+m+i(k_+ - k_-) ]} z^\alpha (1-z)^{-\frac{m}{2}} \frac{\Gamma(c)\Gamma(a_{112}+b_{112}-c)}{\Gamma(a_{112})\Gamma(b_{112}-2)} , 
\end{align}
\begin{align}
R_{122} (z) &\simeq -e_{222}  \frac{m-2-i(k_+ + k_-)}{m-2+i(k_+ - k_-)}  z^\alpha  (1-z)^{-\frac{m}{2}+1} \frac{\Gamma(c)\Gamma(a_{122}+b_{122}-c)}{\Gamma(a_{122})\Gamma(b_{122})} \nn \\
&= - \frac{2e_{222} }{m-2+i(k_+ - k_-)}  z^\alpha  (1-z)^{-\frac{m}{2}+1} \frac{\Gamma(c)\Gamma(a_{122}+b_{122}-c)}{\Gamma(a_{122})\Gamma(b_{122}-1)} ,\\
R_{222} (z) &\simeq e_{222} z^\alpha  (1-z)^{-\frac{m}{2}+2} \frac{\Gamma(c)\Gamma(a_{222}+b_{222}-c)}{\Gamma(a_{222})\Gamma(b_{222})} .
\end{align}
The common set of zeros of the above functions are given by
\begin{eqnarray}
b_{111}-3=b_{112}-2&=&b_{122}-1=b_{222} = -n , \\
a_{111} &=& -n, \qquad n = 0, 1, 2, \ldots, 
\end{eqnarray}
which implies
\begin{equation}
i(k_+ + k_-) = 2n+\hat \Delta -3  , \qquad i(k_+ - k_-) = 2n+\hat \Delta +3 .
\end{equation}
where, $\hat \Delta=1+m$. 
Thus the quasinormal modes for the spin-3 field are,
\begin{eqnarray}
\omega _L &=& k+2\pi T_L(k_+ +k_-), \qquad  \ \omega _R =  -k+2\pi T_R(k_+ -k_-) \\ \nonumber 
				&=&  k-2\pi iT_L(2n + \hat \Delta -3) \qquad 
				=  -k-2\pi iT_R (2n+\hat \Delta +3).
\end{eqnarray}

\section{Proofs of the identities}

In this section we shall be proving the identities (\ref{claim-0}), (\ref{claim-1}) and (\ref{claim-2}). 

\subsection*{Proof of  Identity 1 (\ref{claim-0})}
We require to prove
\begin{equation}
\sum_{q=0}^{s} T^{(s)}_{pq}T^{(s)}_{qr} = 2^s \delta_{pr},
\end{equation}
or equivalently,
\begin{equation}
\sum_{q=0}^{s} \sum_{r=0}^{s} T^{(s)}_{pq}T^{(s)}_{qr} x^r = 2^s \sum_{r=0}^{s} \delta_{pr} x^r = 2^s x^p .
\end{equation}
Let's start with the left hand side
\begin{align}
\sum_{q=0}^{s} \sum_{r=0}^{s} T^{(s)}_{pq}T^{(s)}_{qr} x^r &=\sum_{q=0}^{s} T^{(s)}_{pq} (x+1)^q (x-1)^{s-q} \nn \\
&=(x-1)^s \sum_{q=0}^{s} T^{(s)}_{pq} \left( \frac{x+1}{x-1} \right) ^q \nn \\
&=(x-1)^s \sum_{a=0}^{p} \sum_{b=0}^{s-p}  (-1)^b \binom{p}{a} \binom{s-p}{b}  \left( \frac{x+1}{x-1} \right)  ^{s-a-b} \nn \\
&=(x+1)^s  \sum_{a=0}^{p} \binom{p}{a}   \left( \frac{x-1}{x+1} \right)  ^{a} \ \sum_{b=0}^{s-p} (-1)^b \binom{s-p}{b} \left( \frac{x-1}{x+1} \right)  ^{b} \nn \\
&=(x+1)^s \left( 1+ \frac{x-1}{x+1}\right)^p  \left( 1- \frac{x-1}{x+1}\right)^{s-p} \nn \\
&=(x+1)^s \frac{(2x)^p}{(x+1)^p} \frac{2^{s-p}}{(x+1)^{s-p}} \nn \\
&= 2^s x^p  \qquad \qquad \text{Q.E.D.}
\end{align}
So we have proved
\begin{equation}
T^{(s)}T^{(s)}=2^s \mathbb{I} .
\end{equation}

\subsection*{Proof of Identity 2 (\ref{claim-1})}
We require to prove
\begin{align}
 T^{(s)}M^{(s)}[T^{(s)}]^{-1} &=  D^{(s)} \\
\Rightarrow M^{(s)}[T^{(s)}]^{-1}&=[T^{(s)}]^{-1}D^{(s)} \nn \\
\Rightarrow M^{(s)}T^{(s)}&=T^{(s)}D^{(s)} .
\end{align}
where, $ D^{(s)}_{pq} =-((2p-s)^2+2m(2p-s)+s) \delta_{pq}$.
This is equivalently
\begin{equation}
\label{rel1}
\sum_{b=0}^s \sum_{c=0}^s M^{(s)}_{ab}T^{(s)}_{bc} x^c =\sum_{b=0}^s \sum_{c=0}^s   T^{(s)}_{ab} D^{(s)}_{bc} x^c
\end{equation}
Let's start with the LHS of (\ref{rel1}). Often we shall be using  the variable
\begin{equation}
z=\frac{x+1}{x-1}. 
\end{equation}
Then we have
\begin{align}
&\sum_{b=0}^s  M^{(s)}_{ab}\sum_{c=0}^s T^{(s)}_{bc} x^c \nn \\
=& \sum_{b=0}^s  M^{(s)}_{ab} (x+1)^b (x-1)^{s-b} \nn \\
=& (x-1)^s  \sum_{b=0}^s  M^{(s)}_{ab} z^b \nn \\
=& (x-1)^s  \sum_{b=0}^s [ -(2 s + 2 a (s - a)) \delta _{a,b} -2mp\delta_{a-1,b} -2m(s-a)\delta _{a+1,b} \nn \\
&\qquad\qquad\quad -a(a-1)\delta_{a-2,b} -(s-a)(s-a-1)\delta_{a+2,b} ] z^b \nn \\
=& (x-1)^s  [ -(2 s + 2 a (s - a)) z^a -2ma z^{a-1} -2m(s-a)z^{a+1} \nn \\
&\qquad\qquad\quad -a(a-1)z^{a-2} -(s-a)(s-a-1)z^{a+2} ] \nn \\
=& (x-1)^s z^a [ -(2 s + 2 a (s - a)) -2ma z^{-1} -2m(s-a)z^{1} \nn \\
&\qquad\qquad\quad -a(a-1)z^{-2} -(s-a)(s-a-1)z^{2} ] \nn \\
=& -(x-1)^{s-a-2}(x+1)^{a-2} [ (2 s + 2 a (s - a)) (x-1)^{2}(x+1)^{2}\nn \\
&\qquad\qquad\quad   +2ma(x-1)^{3}(x+1)^{}  +2m(s-a)(x-1)^{}(x+1)^{3}  \nn \\
&\qquad\qquad\quad +a(a-1)(x-1)^{4}  +(s-a)(s-a-1)(x+1)^{4}  ] \nn \\
=& (x+1)^{a-2} (x-1)^{-a+s-2} [-s (x+1)^2 \left( x (-8 a+x-6)+2 m   \left(x^2-1\right)+1 \right) \nn \\
&\qquad\qquad\qquad\qquad\qquad -8 a x \left(2 a x-m x^2+m+x^2+1\right)-s^2 (x+1)^4 ] .
\end{align}
\def\sp{{(s,a)}}
We now need to evaluate the RHS of (\ref{rel1}). Before proceeding we shall list a few definitions and identities which we shall be using the following relations 
\begin{align}
P_{(s,a)}(x) &=\sum_{b=0}^s T^{(s)}_{ab} x^b = (x+1)^{a}(x-1)^{s- a}, \\
x\frac{dP_{(s,a)}(x)}{dx} &= \sum_{b=0}^s b T^{(s)}_{ab} x^b ,\\
x^2 \frac{d^2P_{(s,a)}(x)}{dx^2} &= \sum_{b=0}^s b(b-1) T^{(s)}_{ab} x^b  .
\end{align}
For the RHS of (\ref{rel1})  we have
\begin{align}
&\sum_{b=0}^s T^{(s)}_{ab}\sum_{c=0}^s D^{(s)}_{bc} x^c \nn \\
=& - \sum_{b=0}^s T^{(s)}_{ab}\sum_{c=0}^s [(2b-s)^2 +2m(2b-s)+s]\delta_{bc} x^c \nn \\
=& - \sum_{b=0}^s T^{(s)}_{ab}  [(2b-s)^2 +2m(2b-s)+s] x^b \nn \\
=& - \sum_{b=0}^s T^{(s)}_{ab}  [4b(b-1)+4(1-s+m)b+s(1+s-2m)  ] x^b \nn \\
=& -4\sum_{b=0}^s b(b-1)T^{(s)}_{ab}x^b \ - \ 4(1-s+m)\sum_{b=0}^s b T^{(s)}_{ab}x^b \ - \ s(1+s-2m) \sum_{b=0}^s T^{(s)}_{ab}x^b \nn \\
=& -4x^2\frac{d^2P_\sp(x)}{dx^2} -4(1-s+m)x\frac{dP_\sp(x)}{dx} -s(1+s-2m) P_\sp(x) \nn \\
=& (x+1)^{a-2} (x-1)^{-a+s-2} [-s (x+1)^2 \left( x (-8 a+x-6)+2 m   \left(x^2-1\right)+1 \right) \nn \\
&\qquad\qquad\qquad\qquad\qquad -8 a x \left(2 a x-m x^2+m+x^2+1\right)-s^2 (x+1)^4 ].
\end{align}
This completes the proof of (\ref{claim-1}). 

\subsection*{Proof of Identity 3 (\ref{claim-2})}
\def\s{{(s)}}
We require to prove
\begin{align}
\frac{1}{2^{s-1}} T^{(s-1)} C^{(s)} T^{(s)} &= C'^{(s)} \\
\Rightarrow  C^{(s)} T^{(s)} &= T^{(s-1)}  C'^{(s)} .
\end{align}
where
\begin{align}
C^{(s)}_{a,b} &=(s-a-1)\delta_{a+2,b}+(m-ik_-)\delta_{a+1,b}+(a+ik_+)\delta_{a,b}, \\
C'^{(s)}_{a,b} &=(2a-s+1+m+i(k_+ -k_-))\delta_{a+1,b} \nn \\
& \quad +(2a-s+1+m-i(k_+ +k_-))\delta_{a,b} .
\end{align}
This is equivalent to proving
\begin{equation}
\label{rel2}
\sum _{b=0}^{s} \sum _{c=0}^{s} C^\s _{ab} T^\s _{bc} x^c = \sum _{b=0}^{s-1} \sum _{c=0}^{s} T^{(s-1)} _{ab} C'^\s _{bc}  x^c .
\end{equation}
The LHS of (\ref{rel2}) is
\begin{eqnarray}
\begin{aligned}
& \sum _{b=0}^{s}  C^\s _{ab} \sum _{c=0}^{s} T^\s _{bc} x^c \\
=& \sum _{b=0}^{s}  C^\s _{ab} (x+1)^b (x-1)^{s-b}  \\
=& (x-1)^s  \sum _{b=0}^{s}  C^\s _{ab} z^b   \\
=& (x-1)^s  \sum _{b=0}^{s}[(s-a-1)\delta_{a+2,b}+(m-ik_-)\delta_{a+1,b}+(a+ik_+)\delta_{a,b}] z^b \\
=& (x-1)^s [(s-a-1)z^{a+2}+(m-ik_-)z^{a+1}+(a+ik_+)z^{a}] \\
=& (x-1)^s z^a  [(s-a-1)z^{2}+(m-ik_-)z+(a+ik_+)]  \\
=& (x-1)^{s-a-2} (x+1)^a  [(s-a-1)(x+1)^2 + (m-ik_-)(x+1)(x-1) \\
&\qquad\qquad\qquad\qquad\qquad + (a+ik_+)(x-1)^2] .
\end{aligned}
\end{eqnarray}
Now let's examine the RHS of (\ref{rel2}),
\begin{eqnarray}
\begin{aligned}
&  \sum _{b=0}^{s-1}  T^{(s-1)} _{ab} \sum _{c=0}^{s} C'^\s _{bc} x^c \nn \\
=&  \sum _{b=0}^{s-1}  T^{(s-1)} _{ab} \sum _{c=0}^{s} [ (2b-s+1+m+i(k_+ -k_-))\delta_{b+1,c} \nn \\
&\qquad\qquad\qquad \quad +(2b-s+1+m-i(k_+ +k_-))\delta_{b,c} ]  x^c \nn \\
=&  \sum _{b=0}^{s-1}  T^{(s-1)} _{ab} [ (2b-s+1+m+i(k_+ -k_-))x^{b+1} \nn \\
&\qquad\qquad\qquad \quad +(2b-s+1+m-i(k_+ +k_-))x^{b} ]  \nn \\
=& 2x  \sum _{b=0}^{s-1} b T^{(s-1)} _{ab} x^b + (-s+1+m+i(k_+ -k_-))x \sum _{b=0}^{s-1} T^{(s-1)} _{ab} x^b \nn \\
&+2  \sum _{b=0}^{s-1} b T^{(s-1)} _{ab} x^b + (-s+1+m-i(k_+ +k_-)) \sum _{b=0}^{s-1} T^{(s-1)} _{ab} x^b \nn \\
=& 2x^2 \frac{dP_{(s-1,a)}(x)}{dx} +  (-s+1+m+i(k_+ -k_-))x P_{(s-1,a)}(x) \nn \\
&  +2x \frac{dP_{(s-1,a)}(x)}{dx} +  (-s+1+m-i(k_+ +k_-)) P_{(s-1,a)}(x) \nn \\
=& 2x(x+1) \frac{dP_{(s-1,a)}(x)}{dx} +[(-s+1+m-ik_-)(x+1)+ik_+(x-1)]  P_{(s-1,a)}(x)  \nn \\
=& (x+1)^a (x-1)^{-a+s-2}[ 2 x (-2 a+s-1)-i k_- \left(x^2-1\right)+i k_+ (x-1)^2 \nn \\
& \qquad\qquad\qquad\qquad\qquad +x^2  (m+s-1)-m+s-1]
\end{aligned}
\end{eqnarray}
\begin{align}
=& (x-1)^{s-a-2} (x+1)^a  [(s-a-1)(x+1)^2 + (m-ik_-)(x+1)(x-1)\nn \\
&\qquad\qquad\qquad\qquad\qquad + (a+ik_+)(x-1)^2] .
\end{align}
This completes the proof of (\ref{claim-2}).
\providecommand{\href}[2]{#2}\begingroup\raggedright\endgroup

\end{document}